\documentclass[a4paper,12pt]{article}
\pdfoutput=1
\usepackage{mathrsfs,graphicx,rotating,amsmath,amsfonts,mathtools,booktabs,amssymb,wasysym,bm}
\usepackage[linkcolor={blue},citecolor={red},urlcolor={blue},colorlinks=true,linktocpage=true]{hyperref}
\usepackage{slashed}
\usepackage{subfigure}
\usepackage[nosort]{cite}
\usepackage[table,xcdraw,dvipsnames]{xcolor}
\usepackage{youngtab}
\usepackage{graphicx}
\usepackage{multirow,multicol}
\usepackage{color}

\allowdisplaybreaks

\usepackage[font={small}]{caption}   

\font\tenrsfs=rsfs10 at 12pt
\font\sevenrsfs=rsfs7
\font\fiversfs=rsfs5
\newfam\rsfsfam
\textfont\rsfsfam=\tenrsfs
\scriptfont\rsfsfam=\sevenrsfs
\scriptscriptfont\rsfsfam=\fiversfs
\makeatletter

\newcommand{\bea}{\begin{eqnarray}}
\newcommand{\eea}{\end{eqnarray}}
\newcommand{\SO}{\textrm{SO}}
\newcommand{\SU}{\textrm{SU}}
\newcommand{\U}{\textrm{U}}
\newcommand{\nn}{\nonumber}

\begin{document}

\phantom{a}
\vspace{2cm}

\begin{center}
{\Large\LARGE \bf 
Composite Dynamics in the Early Universe}\\[1cm]

{\bf Stefania De Curtis, Luigi Delle Rose and Giuliano Panico}
\\[7mm]
{\it INFN, Sezione di Firenze, and Department of Physics and Astronomy, \\ University of Florence, Via G. Sansone 1, 50019 Sesto Fiorentino, Italy}\\[5mm]

{\it E-mail:} \texttt{stefania.decurtis@fi.infn.it, luigi.dellerose@fi.infn.it, giuliano.panico@unifi.it}

\vspace{1.5cm}

{\large\bf Abstract}
\begin{quote}

We study the occurrence of a strong first-order electroweak phase transition in composite Higgs models. 
Minimal constructions realising this scenario are based on the coset $\SO(6)/\SO(5)$ which delivers an extended Higgs sector with an additional scalar.
In such models, a two-step phase transition can be obtained with the scalar singlet acquiring a vacuum expectation value at intermediate temperatures.
A bonus of the Nambu--Goldstone boson nature of the scalar-sector dynamics is the presence of non-renormalisable Higgs interactions that can trigger additional sources of CP violation needed to realise baryogenesis at the electroweak scale. Another interesting aspect of this scenario is the generation of gravitational wave signatures that can be observed at future space-based interferometers.

\end{quote}
\thispagestyle{empty}
\bigskip
\end{center}

\newpage
\tableofcontents
\setcounter{footnote}{0}
\newpage

\section{Introduction} 

It is conceivable that the Universe, during its evolution, underwent several spontaneous-symmetry-breaking events.
The study of the corresponding phase transitions is therefore of the utmost theoretical and phenomenological interest.
Among these events, the electroweak (EW) transition plays a privileged role, being responsible for the generation
of the masses of the elementary particles.
An intriguing possibility, that would make this transition even richer, is the fact that it could provide
the necessary conditions to trigger baryogenesis. This scenario is very appealing not only from the theoretical point
of view but also from the experimental perspective. Apart from the possible cosmological signatures, the dynamics
involved in EW baryogenesis (EWBG) is tied to the TeV energy scale, and is therefore testable
at present and near-future collider experiments~\cite{Curtin:2014jma}. This has to be contrasted, for instance, with the
vanilla leptogenesis scenario in which the scale of new physics is typically $\Lambda_\textrm{NP} \gtrsim 10^9\,{\rm GeV}$,
well above the reach of foreseeable collider probes.

Realising EWBG, however, is not easy. As well-known, the Standard Model (SM) fails in satisfying quantitatively
the Sakharov conditions.
In fact, CP violation from the CKM matrix is not enough to guarantee the generation of the observed
matter-antimatter asymmetry and the EW phase transition (EWPhT) it is not sufficiently strong
(it is actually only a cross-over) to trigger a significant departure from thermal equilibrium.
As a consequence, baryogenesis at the EW scale necessarily requires new physics Beyond the SM (BSM), in particular
new sources of CP violation and a modified (possibly enlarged) scalar sector that could provide a sufficiently strong
first-order EWPhT.

Another interesting aspect of EWBG is provided by the additional observational consequences related to
gravitational waves (GWs). Indeed, the violent environment due to a strong first-order phase transition gives rise to a significant
amount of energy released in the form of GWs.
The peak frequency of this signal is unfortunately too low to be tested at present
ground-based interferometers. However near-future space-based experiments, such as LISA, BBO and DECIGO,
will be sensitive to the interesting range of wave spectra.

The possibility to achieve a first order phase transition at the EW scale has been scrutinised in several BSM scenarios. 
For example, in the minimal supersymmetric SM a sufficiently light stop squark can provide a large contribution, through thermal loops, to the cubic term in the temperature-dependent effective potential~\cite{Carena:1996wj,Delepine:1996vn}.
Alternatively, a barrier can be generated already at tree-level
by an extended scalar sector as in the singlet-augmented SM~\cite{Espinosa:2011ax,Cline:2012hg,Profumo:2014opa,Curtin:2014jma,Kakizaki:2015wua,Vaskonen:2016yiu,Kurup:2017dzf} or in 2-Higgs doublet models~\cite{Fromme:2006cm,Dorsch:2013wja,Dorsch:2014qja,Dorsch:2016nrg}.
Recently, some attention has also been devoted to the cosmological consequences of a dilaton state emerging from the spontaneous breaking of the conformal symmetry of a strongly coupled dynamics~\cite{Konstandin:2011dr,Bruggisser:2018mus,Bruggisser:2018mrt,Baratella:2018pxi}.
Finally model-independent approaches based on the effective-field theory parametrisation have been considered in refs.~\cite{Bodeker:2004ws,Grojean:2004xa,Delaunay:2007wb,Grinstein:2008qi}, in which the impact of higher-order terms
in the Higgs potential has been studied.

New physics models in which the Higgs boson arises as a pseudo Nambu--Goldstone Boson (NGB)
from a strongly interacting theory can provide other natural scenarios in which first-order transitions can be obtained.
The simplest models realising the idea of a NGB Higgs, the minimal Composite Higgs Models (CHMs)~\cite{Agashe:2004rs},
are based on the symmetry breaking pattern $\SO(5) \rightarrow \SO(4)$, which delivers the four real scalar degrees
of freedom that build-up a SM-like $\SU(2)$ Higgs doublet.
In these scenarios the Higgs couplings are modified, but the strong experimental constraints already put an upper bound
of order $10\%$ on these deviations~\cite{Grojean:2013qca,Aad:2015pla}. Moreover direct and indirect searches exclude
additional composite resonances for masses below $\sim 1\,{\rm TeV}$~\cite{Matsedonskyi:2015dns,Sirunyan:2018yun,Aad:2019fbh}. For these reasons the corrections to the
Higgs thermal potential are necessarily small and can not convert the EWPhT into a strong first-order one.

Successful CHMs exhibiting a first order EW transition can instead be realised by considering non-minimal symmetry-breaking
patterns. A simple possibility, on which we will focus in this work,
is provided by the coset $\SO(6)/\SO(5)$~\cite{Gripaios:2009pe,Frigerio:2012uc}, in which
the Higgs doublet is accompanied by an extra real scalar $\eta$\footnote{Additional composite Higgs models with larger cosets have been considered in refs.~\cite{Chala:2016ykx,Chala:2018opy}}. In this case the effective potential for the scalar fields
allows for a much richer phase-transition behaviour.
The EW transition may proceed directly from the symmetric phase to the EW-symmetry-breaking (EWSB) one
or via some intermediate steps. The latter possibility is particularly interesting. As we will see, in a large part
of the parameter space of the model, a two-step EWSB transition can be realised, in which
the singlet gets a vacuum expectation value (VEV) at intermediate temperatures.
In such case, the presence of a $H^2 \eta^2$ portal interaction creates a barrier between the EW vacuum and
the intermediate vacuum, giving rise to a first-order transition that can be strong enough to allow for baryogenesis.

Another interesting aspect of the composite scenario is the fact that the NGB nature of the Higgs implies the presence
of non-renormalisable Higgs interactions that can provide additional sources of CP violation. As we will see,
interactions of the form $\eta\,h\,\bar t_L\,t_R$, are naturally there in $\SO(6)/\SO(5)$ models and can trigger
CP-violating effects during the nucleation of bubbles in the first-order transition.

The paper is organised as follows. In sec.~\ref{sec:ewphtSO5} we briefly review the properties of the EWPhT
in minimal CHMs based on the $\SO(5)/\SO(4)$ coset. In sec.~\ref{sec:ewpht} we study the general structure of the
effective potential for a scenario with an extra singlet scalar. In particular we consider a renormalisable potential,
which, as we will see, can be used as a good approximation for the full potential in the NGB Higgs case.
By exploiting a high-temperature approximation, we find the conditions that allow for a two-step EW transition and
derive approximate analytical expressions for the position of the minima and the phase transition temperature.
In sec.~\ref{sec:coset} we consider the $\SO(6)/\SO(5)$ composite Higgs scenarios.
We inspect several embeddings for the SM fermions, matching the full effective potential with the
renormalisable one introduced in sec.~\ref{sec:ewpht}. In particular, we discuss the range of parameters accessible
in each explicit scenario and the generation of the $\eta\, h\,\bar t_L\, t_R$ operator.
In sec.~\ref{sec:paramspace} we explore the properties of the EWPhT. We determine the regions of parameter
space in which a two-step transition can happen. We also determine numerically some important quantities that
characterise the first-order transition, namely the critical and nucleation temperatures, the strength of the transition,
the vacuum energy density, the width of the bubble wall and the inverse duration of the transition. All these parameters
are important because they control the generation of GWs and the possibility to achieve baryogenesis.
The spectrum of the GWs and the possibility to test them at future space-based interferometers are discussed in
sec.~\ref{sec:gw}, while the possibility to achieve baryogenesis is investigated in sec.~\ref{sec:EW_baryo}. In the latter
section we also discuss the generation of CP violation and the bounds coming from the experimental flavour data.
Finally in sec.~\ref{sec:conclusions} we present our conclusions.

In the appendices we collect additional results on the effective potential in $\SO(6)/\SO(5)$ CHMs (appendix~~\ref{app:potential}) and analytical results for the properties of the EWPhT obtained in the thin-wall
approximation (appendix~\ref{app:thinwall}).

\section{Models with a minimal Higgs sector}
\label{sec:ewphtSO5}

Before considering the class of scenarios we are interested in, namely the ones with a Higgs sector extended with an additional
singlet, we briefly review the properties of the EWPhT in minimal CHMs
with only one Higgs doublet. This preliminary discussion is useful to fix our notation and discuss a few key approximations
that will be employed in the main analysis.

The simplest viable CHM is realised through the coset $\SO(5)/\SO(4)$~\cite{Agashe:2004rs}. This symmetry pattern gives rise to
one Higgs doublet and preserves a custodial invariance, which helps in keeping under control dangerous corrections to the
precision EW parameters.

The leading terms in the scalar potential for the physical Higgs field $h$ can be schematically written as
\bea
\label{eq:MCHMpot}
V(h) = - \alpha f^2 \sin^2 \frac{h}{f} + \beta f^2 \sin^4 \frac{h}{f} \,,
\eea
where $f$ is the Goldstone Higgs decay constant.
The main contributions to the parameters $\alpha, \beta$ come from the top and fermionic top-partner sector,
whereas the gauge sector typically gives smaller corrections.\footnote{In eq.~(\ref{eq:MCHMpot})
we assumed that the top partners transform in the fundamental representation of $\SO(5)$.
For different representations the periodicity of the potential is in general different. For our illustrative purposes this detail
is however irrelevant.}
Regardless of the details of the model, the parameters $\alpha, \beta$ must reproduce a realistic EWSB and the correct Higgs mass. As such, they are forced to satisfy the tuning conditions
$\alpha = 2 \beta \xi$ and $m_h^2 = 8 \xi (1 - \xi) \beta$ with $\xi = \sin^2 \langle h \rangle/f$.
A minimal amount of tuning of order $1/\xi$ is therefore unavoidable~\cite{Panico:2012uw}.

The one-loop thermal corrections to the Higgs potential are given by
\bea
V_{T}(h,T) = \sum_{b} \frac{n_b T^4}{2 \pi^2} J_B\left( \frac{m_b^2(h)}{T^2} \right) - \sum_{f} \frac{n_f T^4}{2 \pi^2} J_F\left( \frac{m_f^2(h)}{T^2} \right)
\eea
where $b,f$ run over the bosonic and fermionic degrees of freedom, with $n_b$ and $n_f$ the corresponding number of
degrees of freedom. The thermal integrals are
\bea
J_B(y) =  \int_0^\infty dx \, x^2 \log \left[ 1 - e^{- \sqrt{x^2 + y}} \right], \qquad
J_F(y) =   \int_0^\infty dx \, x^2 \log \left[ 1 + e^{- \sqrt{x^2 + y}} \right] \,.
\eea
We are interested in temperatures (much) below the mass scale of the composite resonances, so the only fields that
give rise to relevant thermal corrections are the SM degrees of freedom, in particular the gauge fields, the top quark
and the Higgs itself. The masses of the SM fields are given by
\bea
& \displaystyle m_W^2(h) = \frac{g^2}{4} f^2 \sin^2 \frac{h}{f} \,, \quad m_Z^2(h) = \frac{g^2 + g'^2}{4} f^2 \sin^2 \frac{h}{f} \,,\nn\\
& \displaystyle m_t^2(h) = \frac{y_t^2}{2} f^2 \sin^2 \frac{h}{f} \,, \nn \\
& \displaystyle m_h^2(h) = \frac{m_h^2}{4 \xi (\xi - 1)} \left( (2\xi -1) \cos 2\frac{h}{f}  + \cos 4 \frac{h}{f}\right) \,.
\eea
In order to tame the infrared singularity of the finite-temperature effective potential in the high $T$ limit, we also implemented the one-loop ring-improved corrections by replacing the $T = 0$ masses of the scalars
and of the gauge bosons with the corresponding thermal masses.

\begin{figure}
\centering
\includegraphics[width=0.47\textwidth]{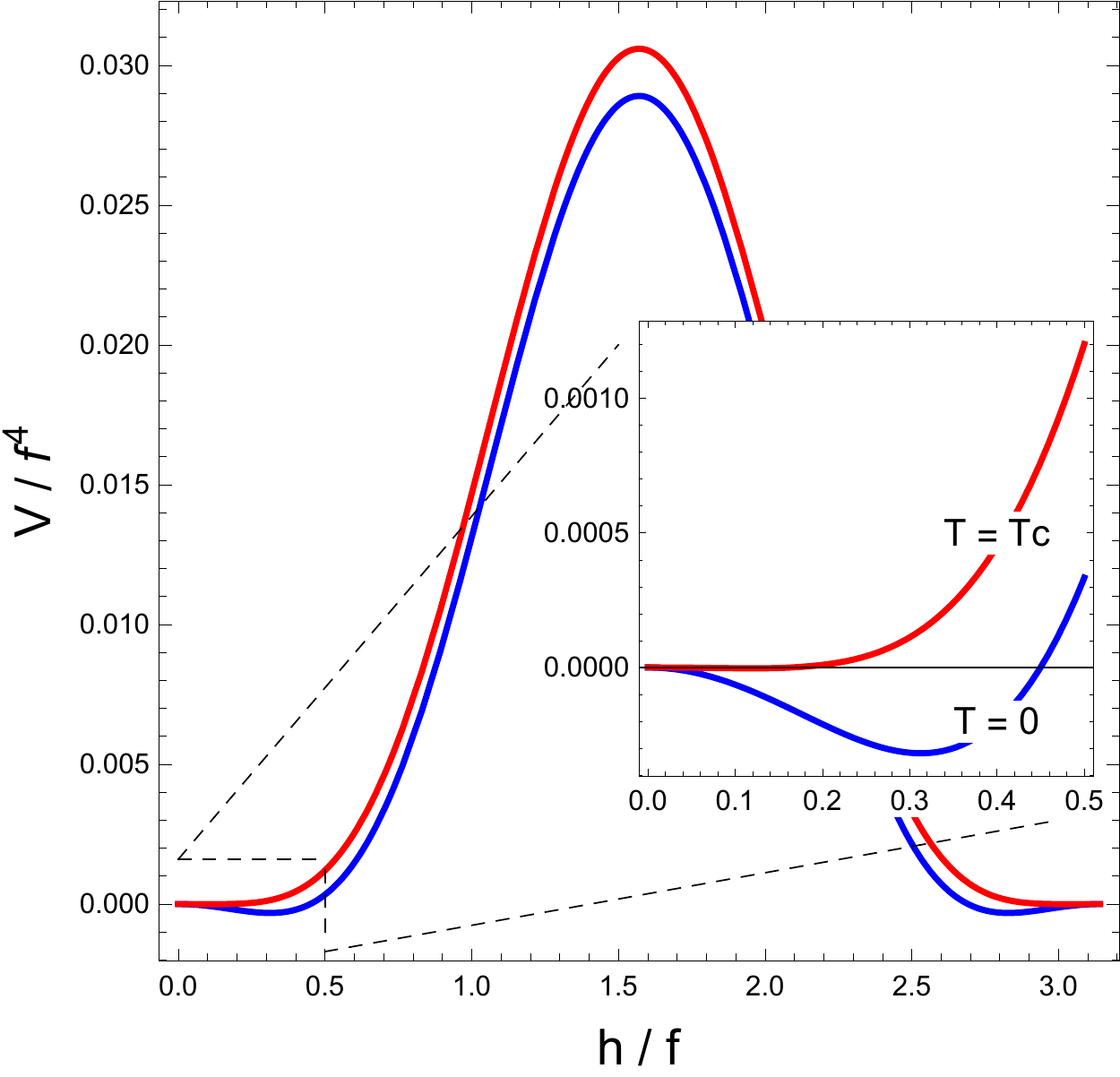} \hfill
\raisebox{.52em}{\includegraphics[width=0.44\textwidth]{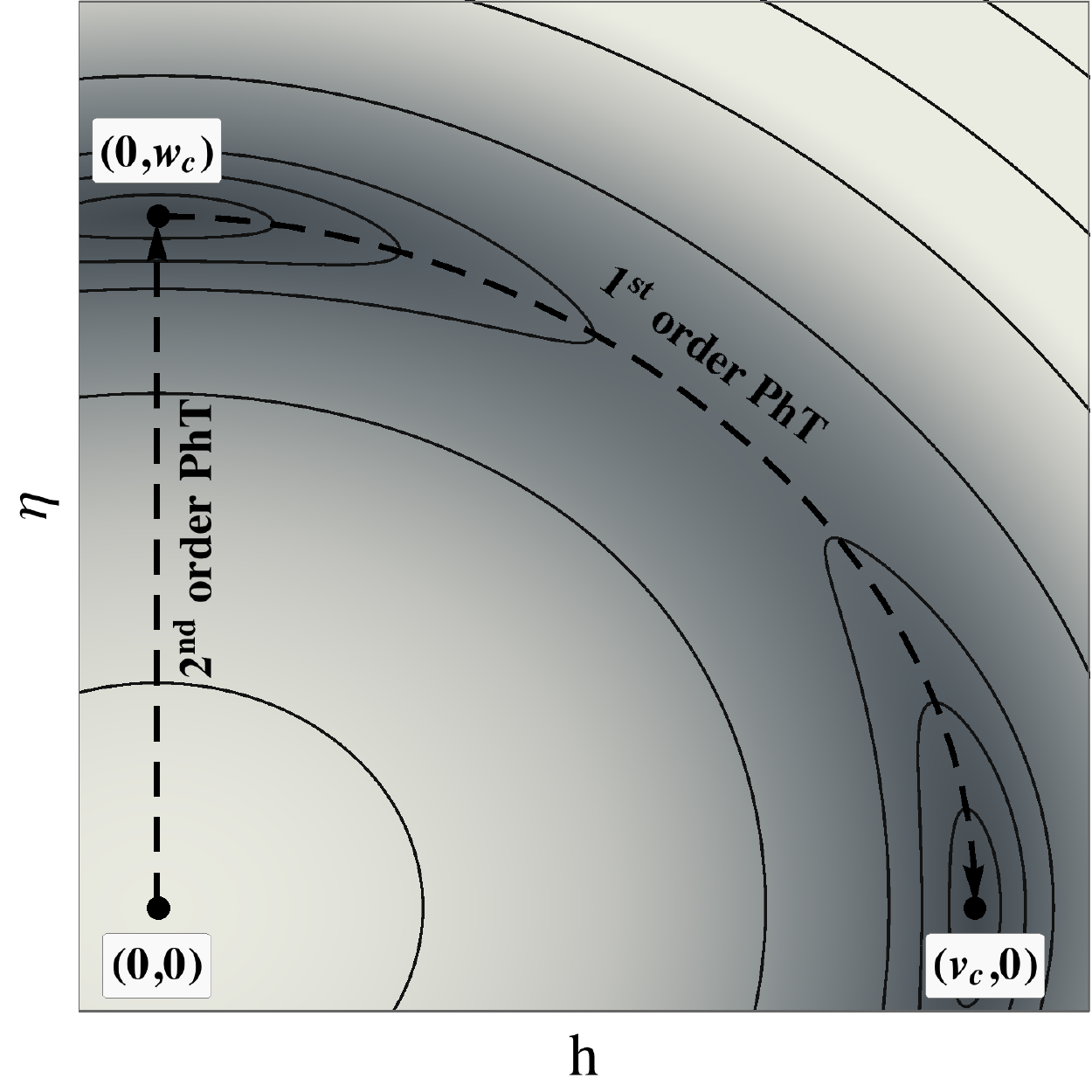}}
\caption{Left panel: normalised scalar potential in the $\SO(5)/\SO(4)$ CHM at $T = 0$ (blue curve) and at the critical temperature (red curve). The parameters are chosen to correctly reproduce the EW vacuum and the Higgs mass. Right panel: schematic illustration of the two-step phase transition. A darker colour corresponds to a deeper potential at the critical temperature $T_c$.}
\label{fig:fig1}
\end{figure}

The structure of the potential and the thermal corrections are analogous to the SM ones. The only difference comes from the
presence of non-linearities due to the Goldstone nature of the Higgs, whose size is controlled by the $\xi$ parameter.
The current experimental bounds, from EW precision measurements~\cite{Grojean:2013qca}, Higgs coupling measurements~\cite{Aad:2015pla}
and direct searches for top partners~\cite{Matsedonskyi:2014mna,Matsedonskyi:2015dns}, impose the quite stringent constraint
$\xi \lesssim 0.1$ on the compositeness scale.
The effects of the non-linearities are therefore small and do not significantly modify the properties of the EWPhT. 
As a consequence we expect, for a heavy enough Higgs ($m_h \gtrsim 80\;\rm{GeV}$), the EWPhT to be
rather weak, possibly just a cross over, as in the SM. This expectation is confirmed by a numerical analysis (see left panel
of fig.~\ref{fig:fig1}) which shows that no significant barrier is present at the critical temperature.\footnote{As shown in ref.~\cite{DiLuzio:2019wsw}, the degeneracy between the $h = 0$ and $h = \pi f$ vacua can be lifted by a small tilt in the potential. This allows for quantum tunnelling through a barrier from the metastable vacuum to the true ground state if,
during the cosmological history, the system ended up in the false vacuum configuration.}

As we already mentioned the validity of the previous discussion is restricted to temperatures below the mass of the
composite resonances, $T \ll m_{\textit lightest}$, with $m_{\textit lightest}$ being the mass of the lightest resonance. As the temperature approaches the mass of the composite states, the compositeness scale drops rapidly to zero~\cite{Barducci:1987gn,Barducci:1989eu}. 
For higher temperatures the global symmetry of the strongly interacting theory is restored and the description of the light degrees of freedom in terms of the chiral Lagrangian is no longer correct.
In minimal models the masses of the composite resonances are strongly constrained by the direct experimental searches.
In particular the fermionic top partners are excluded up to masses of order $1\;{\rm TeV}$ (see for instance ref.~\cite{Sirunyan:2018yun}) and the vector resonances
up to $\sim 2-3\;{\rm TeV}$ (see for instance ref.~\cite{Aad:2019fbh}). The EWPhT in these models happens at temperatures well below the TeV scale,
so that the approximation of neglecting the resonances effects is fully justified.\footnote{The effects of resonances on
the properties of the EWPhT have been studied within the holographic realizations of the composite Higgs
scenarios~\cite{Panico:2005ft}. The results confirm that resonances with a mass larger than the critical temperature
have a small impact on the properties of the phase transition.}

\section{Models with an additional singlet}
\label{sec:ewpht}   

The failure of the minimal CHM in realising a first order EWPhT motivates the exploration of more complex scenarios with
extended global symmetries and a non-minimal Higgs sector.
In this respect, CHMs based on the $\SO(6)/\SO(5)$ coset~\cite{Gripaios:2009pe} are very promising since they predict an extra scalar,
neutral under the SM group. As well-known, in the elementary singlet-extended SM,
the presence of a light scalar can help achieving a first order phase transition through a tree-level barrier
in the scalar potential.

As we saw in the previous section, phenomenologically viable models require the VEV
of the scalar fields in the Higgs
sector to be significantly smaller than the compositeness scale $f$, so that $\xi \ll 1$. In this regime, the non-linearities
due to the Goldstone nature of the Higgs are small and the whole potential can be well approximated by a simple expansion
including quadratic and quartic terms in the scalar fields. This approximation allows us to map the potential onto the
one of the elementary singlet-extended SM.
For phenomenological reasons that we will discuss later on, we are interested in models characterised by a
(possibly approximate) discrete $Z_2$ symmetry $P_\eta$ under which the additional scalar $\eta$ switches sign.
In this case the zero-temperature scalar potential for the physical Higgs $h$ and the $\eta$ singlet can be parametrised
up to quartic terms as
\bea
\label{eq:V_elem}
V(h, \eta) =  \frac{\mu_h^2}{2} h^2 + \frac{\lambda_h}{4} h^4 + \frac{\mu_\eta^2}{2} \eta^2 + \frac{\lambda_\eta}{4} \eta^4 + \frac{\lambda_{h \eta}}{2} h^2 \eta^2\,.
\eea

We can understand the main features of the EWPhT in this model by studying the thermal potential in the leading
high-temperature expansion. In this approximation the scalar fields acquire temperature-dependent masses given by
\bea
\label{eq:V_highT}
V_T(h, \eta, T) =  \left( c_h \frac{h^2}{2} + c_\eta \frac{\eta^2}{2} \right) T^2\,.
\eea
The coefficients $c_h$ and $c_\eta$, taking into account the corrections coming from the gauge bosons,
the top and the scalar sector, are given by
\bea
\label{eq:ch_ceta}
c_h = \frac{1}{48} \left( 9 g^2 + 3 g'^2 + 12 y_t^2 + 24 \lambda_h + 2 \lambda_{h \eta}   \right) \,, \qquad
c_\eta = \frac{1}{12} \left( 4 \lambda_{h\eta}  + \lambda_{\eta}\right) \,.
\eea
At very large temperatures, the thermal masses bring the vacuum in the symmetric configuration $(h,\eta) = (0,0)$. 
When the temperature drops down, the symmetric minimum becomes unstable and the system eventually ends into
the EWSB vacuum $(v = 246\;{\rm GeV}, 0)$ at $T = 0$. 
This is obviously possible only if the $(v,0)$ configuration is the global minimum of the potential, which is ensured by the condition 
\bea
\label{eq:lambdaeta}
\lambda_\eta > \frac{\mu_\eta^4}{\mu_h^4}\lambda_h = 2 \frac{(m_\eta^2 - \lambda_{h \eta} v^2)^2}{m_h^2 v^2} \,,
\eea
requiring a sufficiently large quartic coupling for the singlet.

The transition to the electroweak vacuum may happen in two ways depending on the parameters of the potential:
through a single, usually dubbed ``one-step'', transition, or through a ``two-step'' process.
In the first case, the singlet never gets a VEV during the cosmological history, $\mu_\eta > 0$. A first-order phase transition
can be triggered by zero-temperature loop effects, but this can happen only for very large values of the portal coupling
$\lambda_{h\eta}$, which, as we will see, are difficult to get in CHMs.
For smaller values of $\lambda_{h\eta}$ a second order transition (or a cross over) is obtained.
In the two-step case, instead,
the singlet $\eta$ temporarily acquires a non-zero VEV at intermediate temperatures and only at a lower temperature a transition to the EWSB vacuum is realised. This process is schematically shown in the right panel of fig.~\ref{fig:fig1}. We will focus our discussion on the latter scenario, in particular on the case in which the zero-temperature
minimum is the EWSB one $(v,0)$. This scenario requires the following two conditions
\begin{equation}\label{eq:conditions_eta}
\mu_\eta^2 < 0\,, \qquad \quad \mu_\eta^2 + \lambda_{h\eta} v^2 > 0\,.
\end{equation}
The first one is obviously needed to trigger a VEV for the singlet at high temperature, while the second one
ensures that the EWSB vacuum is the global minimum of the potential at zero temperature.

The two-step pattern can be easily achieved since typically $c_h > c_\eta$, due to the fact that the Higgs field couples
to a larger set of light degrees of freedom. Since the zero-temperature mass terms $\mu_h$ and $\mu_\eta$ are usually
of the same order, the $\eta$ field can get a negative squared mass at higher temperatures than the Higgs,
thus triggering a VEV for $\eta$ but no breaking of the EW symmetry. At lower temperatures the potential develops
an additional minimum along the $h$ direction, generating two minima separated by a tree-level barrier.
If the EWSB minimum $(v,0)$ is the global one at sufficiently low temperatures, when the Universe cools down,
the system collapses in the true ground state with a first-order phase transition.
This process is described by the nucleation of bubbles of true vacuum in a background of the unstable EW-symmetric configuration. The strength of the transition is parametrised by $v_c/T_c$.
This is a crucial parameter controlling the efficiency of EWBG and, in particular, a strong first-order transition with $v_c/T_c > 0.6 - 1.6$ \cite{Patel:2011th,Curtin:2014jma} prevents sphalerons to wash out the baryon asymmetry inside the broken phase.

The critical temperature $T_c$ and the corresponding position of the minima $(v_c, 0)$ and $(0, w_c)$ can be easily
computed in the high-temperature expansion and are given by
\bea
v_c^2 = \frac{c_\eta \, \mu_h^2 - c_h \, \mu_\eta^2}{c_h \sqrt{\lambda_h \lambda_\eta} - c_\eta \, \lambda_h} \,, \qquad w_c^2 = v_c^2  \sqrt{\frac{\lambda_h}{\lambda_\eta}} \,, \qquad 
\frac{v_c^2}{T_c^2} = \frac{c_\eta \, \mu_h^2 - c_h \, \mu_\eta^2}{\lambda_h \, \mu_\eta^2 - \sqrt{\lambda_h \lambda_\eta} \, \mu_h^2} \,.
\eea   
In order to ensure that the $(0, w_c)$ minimum becomes deeper than the EWSB one at intermediate temperatures,
the following conditions must be satisfied
\begin{equation}
c_h/c_\eta >  \sqrt{\lambda_h/\lambda_\eta}\,,
\qquad \quad
c_h/c_\eta >  {\mu_h^2/\mu_\eta^2}\,.
\end{equation}

\section{The $\SO(6)/\SO(5)$ model}
\label{sec:coset}

We can now focus on the class of models involving a strong sector characterised by the symmetry breaking pattern
$\SO(6) \to \SO(5)$. This symmetry structure delivers five NGB fields transforming as the $\bf 5$ of $\SO(5)$
and decomposing as $\bf 4 \oplus \bf 1 \simeq (\bf 2, \bf 2) \oplus (\bf 1 , \bf 1)$ under the subgroup
$\SO(4) \simeq \SU(2)_L \times \SU(2)_R$. As usual, the bidoublet describes the SM Higgs fields
while the singlet component corresponds to an extra scalar degree of freedom
in the NGB spectrum.

The NGB fields $\pi^{\hat a}$ in the fundamental representation of $\SO(6)$ are described by the matrix
\begin{equation}
U = \exp \left( i \frac{\Pi}{f} \right)  =  1 + i \frac{s_\pi}{\pi} \Pi + \frac{c_\pi - 1}{\pi^2} \Pi^2
\quad \textrm{with} \quad 
\Pi = \sqrt{2} \pi_{\hat a} T^{\hat a} = - i \left( \begin{array}{cc}
0_{5 \times 5} & \bm {\pi} \\
- {\bm \pi}^T & 0
\end{array} \right)\,,
\end{equation}
where the index $\hat a$ runs over the five broken $\SO(6)$ generators 
$T^{\hat a}_{ij} = - \frac{i}{\sqrt{2}} \left( \delta^{\hat a}_i  \delta^{\hat 6}_j  -  \delta^{\hat a}_j  \delta^{\hat 6}_i \right)$ with $\hat a = 1 , \ldots, 5$,
and we used the short-hand notation 
$s_\pi = \sin \frac{\pi}{f}$, $c_\pi = \cos \frac{\pi}{f}$ and $\pi = \sqrt{\pi_{\hat a} \pi_{\hat a}}$.

In order to describe the dynamics of the Goldstones and derive the expression of the effective potential, it is
useful to choose a reference direction for the spontaneous breaking of the $\SO(6)$ symmetry, namely
$\Sigma_0 = (0,0,0,0,0,1)^T$. The Goldstones can be thus encoded in the multiplet
\bea
\Sigma &=& U \Sigma_0 = \left( \frac{\pi_i}{\pi} s_\pi, c_\pi \right)^T
=   \left( \frac{h_i}{h} s_h , s_\eta ,
\sqrt{1 - s_h^2 - s_\eta^2}\right)^T\,.\label{eq:Gold_param}
\eea
Notice that in the second equality we introduced a different parametrisation of the Goldstones, in which $h_i$ ($i = 1,2,3,4$), describe the Higgs doublet components and $\eta$ the singlet ($s_h$ and $s_\eta$ are defined in analogy with $s_\pi$).
This parametrisation is more convenient for small values of the Higgs and $\eta$ VEVs, as the ones we will be interested in
our analysis. This redefinition, indeed, ensures that the scalar fields are canonically normalized in all configurations
in which either $\langle h \rangle = 0$ or $\langle \eta \rangle = 0$. This can be seen from the following
Lagrangian including the kinetic terms for the physical Higgs $h$ and the $\eta$ as well as the interactions
with the gauge fields
\bea
\mathcal L_{kin} = \frac{f^2}{2} |\mathcal D_\mu \Sigma |^2 &=& \frac{1}{2} \bigg[ (\partial_\mu h)^2   +   (\partial_\mu \eta)^2 + \frac{1}{f^2} h\, \eta\, \partial_\mu h\, \partial^\mu \eta  + \cdots \nn \\
&& + g^2 f^2 \sin^2 h/f \left( W^{\mu+} W^-_\mu + \frac{1}{2 \cos^2 \theta_W} Z^\mu Z_\mu \right) \bigg]\,,
\eea
where the dots on the first line denote additional higher-order terms involving derivatives of the scalar fields.

It is important to stress that the non-linear Goldstone structure unavoidably leads to mixed kinetic interactions involving both $h$ and $\eta$.
In the form we choose, however, these terms are not relevant in the minima of the potential (in which at least one of the
scalars has vanishing VEV). Possible effects are instead present away from these minima, in particular they can determine a
distortion of the bounce trajectory that the fields follow during the phase transition from the $(0, w_c)$ to the EWSB
vacuum $(v_c, 0)$. We checked numerically that these effects are small (at most at the percent level), so that
they can be safely neglected in the numerical studies.

Notice that the parametrisation in eq.~(\ref{eq:Gold_param}) reduces to the usual $\SO(5)_H \to \SO(4)$ symmetry
breaking structure for $\eta = 0$, where $\SO(5)_H$ is identified with the $\SO(6)$ rotations that leave invariant
the $5$-th coordinate.
On the other hand, for $h_i = 0$, one obtains an $\SO(2)_\eta \rightarrow {\textit nothing}$ non-linear $\sigma$-model,
where $\SO(2)_\eta$ corresponds to the rotations in the $5$-th and $6$-th coordinates.
These limits will be useful later on to understand the structure of the effective potential
for the Higgs and the additional scalar singlet.

Before concluding our general discussion of the $\SO(6)/\SO(5)$ structure, we consider the discrete symmetries that
can characterise the composite sector and its mixing with the elementary states. In particular we highlight the discrete $Z_2$
symmetry $P_\eta$, under which~\cite{Frigerio:2012uc}
\begin{equation}
(h_1, h_2, h_3, h_4 , \eta) \rightarrow (h_1, h_2, h_3, h_4, -\eta)\,.
\end{equation}
This symmetry is not part of $\SO(5)$ (nor of $\SO(6)$), but commutes with $\SO(4)$ and can be seen as a parity transformation extending $\SO(6)/\SO(5)$ to ${\rm O}(6)/{\rm O}(5)$.
Under $P_\eta$ the Goldstone matrix $U$ transforms as $U \rightarrow {P_\eta}\cdot U\cdot{P_\eta}$, with
\begin{equation}
{P}_\eta = {\rm diag}(1, 1, 1, 1, -1, 1)\,.
\end{equation}
The $\sigma$-model Lagrangian at the two-derivative level is automatically endowed with the ${P}_\eta$ symmetry,
while higher-order terms do not necessarily preserve it (for instance operators involving the Levi--Civita tensor, such as the Wess--Zumino--Witten term).
As we will see in the following, ${P}_\eta$ is also useful to forbid a tadpole for $\eta$ in the fermionic contribution to the effective potential, leading to the $Z_2$ symmetric form given in eq.~(\ref{eq:V_elem}).

\subsection{The fermion sector}

Under the framework of partial compositeness~\cite{Kaplan:1991dc}, the elementary fermions couple linearly to operators of the strong sector, $\lambda_\psi \bar \psi\, \mathcal O_\psi$, 
whose quantum numbers determine the SM fermion interactions and their contributions to the effective potential of the NGB fields. 
The latter is generated by the explicit breaking of $\SO(6)$ since the elementary fermions do not fill complete representations of the global symmetry group. 
In the following, we will focus our attention on the top sector since it gives, in general, the largest contribution among the SM fermions. When necessary, we will introduce the lighter quarks and comment on their impact.

Several representations for the composite fermionic operators can be considered.
The simplest ones, and their decomposition under $\SO(4) \times \SO(2)_\eta \simeq \SU(2)_L \times \SU(2)_R \times \U(1)_\eta$~\cite{Frigerio:2012uc}, are given by
\bea
\bold 1 &=& (\bold 1, \bold 1)_{0} \,, \nn \\
\bold 4   &=& (\bold 2, \bold 1)_{+1} \oplus (\bold 1, \bold 2)_{-1} \,, \nn \\
\bold 6   &=& (\bold 2, \bold 2)_{0} \oplus (\bold 1, \bold 1)_{+2} \oplus (\bold 1, \bold 1)_{-2} \,, \nn \\
\bold{10} &=& (\bold 2, \bold 2)_{0} \oplus (\bold 3, \bold 1)_{+2} \oplus (\bold 1, \bold 3)_{-2} \,, \nn \\
\bold{15} &=& (\bold 1, \bold 3)_{0} \oplus (\bold 3, \bold 1)_{0} \oplus (\bold 1, \bold 1)_{0} \oplus  (\bold 2, \bold 2)_{+2} \oplus  (\bold 2, \bold 2)_{-2} \,, \nn \\
\bold{20}' &=& (\bold 3, \bold 3)_{0} \oplus  (\bold 2, \bold 2)_{+2} \oplus  (\bold 2, \bold 2)_{-2}  \oplus (\bold 1, \bold 1)_{+4} \oplus (\bold 1, \bold 1)_{-4} \oplus (\bold 1, \bold 1)_{0}
\label{eq:reps}
\eea
The singlet representation $\bold 1$ can obviously be used only to embed the right-handed quark components and must
be necessarily complemented by another representation for the left-handed quark doublets. 
If the latter are embedded in the $\bold 6$, a potential for the scalar singlet is not developed, while if the $\bold{15}$ is chosen, a mass for the top quark cannot be generated. 
Instead, if the left-handed quark doublets are embedded in the $\bold{20}'$ with right-handed quarks in the $\bold 1$, the scenario that emerges is not dissimilar from the simpler one with both quarks in the fundamental representation. For these reasons we will not consider models with the right-handed quarks in the $\bold 1$ representation.
 
The spinor representation $\bold 4$ does not contain a $(\bold 2, \bold 2)$ and, as such, is not adequate for the embedding of the top quark doublet since it would generated a large correction to the $Z \bar b_L b_L$ coupling.

The $\bold{10}$ is suitable for the embedding of the left-handed quark doublet in the $(\bold 2, \bold 2)_{0}$ and the right-handed quark in the $(\bold 1, \bold 3)_{-2}$. Nevertheless, these embeddings would not generate a potential for the scalar singlet since they do not break $\SO(2)_\eta$. 

For these considerations we will focus our discussion on the properties of the representations $\bold 6, \bold{15}$ and $\bold{20}'$.

From the collider viewpoint, these alternative scenarios could be distinguished by looking at the spectrum of top (and possibly bottom) partners.
Each embedding gives rise to multiplets of fermionic resonances with specific electroweak quantum numbers, which can be typically determined
by analysing the particle decay modes. For the main $\textrm{SO}(6)$ representations, the phenomenology of top partners has been already scrutinised in the literature. The signatures of extended multiplets (such as the $\bf 20'$) are still partially unexplored. Such analysis is however beyond the scope of our paper.

\subsubsection{Fermions in the fundamental representation}
\label{sec:(6,6)-model}

The simplest model is realised by embedding the quark doublet $q_L$ and the right-handed components $u_R, d_R$ in the fundamental representation of $\SO(6)$.  We denote this set-up as the $({\bf 6}, {\bf 6})$ model.
As well known, in order to accommodate the correct hypercharges for the fermions, an additional unbroken
$\U(1)_X$ global symmetry must be introduced in the composite sector (see appendix~\ref{app:gauge} for more
details).
The $\U(1)_X$ charges of the right-handed fields are, respectively, $X_{u_R} = 2/3$ and $X_{d_R} = -1/3$, required to reproduce the correct hypercharge assignment. 
Consequently, the $q_L$ must be embedded into two multiplets with different $\U(1)_X$ charge, namely one with $X_{q_L} = 2/3$ in order to generate a mass term for the up-type quarks, and a second one with $X_{q_L'} = -1/3$
for the mass term of the down-type quarks.\footnote{Since the $(\bold 2, \bold 2)_{X=-1/3}$ breaks the custodial symmetry that protects the $Z \bar b_L b_L$ coupling from large corrections,
it is important to assume that the quark doublet is mostly described by the $(\bold 2, \bold 2)_{X=2/3}$ representation which is usually guaranteed by the smallness of the mass of the bottom quark with respect to that of the top.}
Notice also that the $q_L$ embedding does not break $\SO(2)_\eta$ and does not generate a potential for the singlet $\eta$.

In order to derive the form of the effective potential, it is useful to uplift the mixing of the elementary fermions
to the composite operators to a formally $\SO(6)$-invariant form. This can be done by the introduction of spurions,
namely~\cite{Panico:2011pw,DeSimone:2012fs,Panico:2015jxa}
\begin{equation}
{\cal L}_{\textit ferm} = \bar q_L^\alpha (\Lambda_{q_L})_\alpha^I ({\cal O}_{q_L})_I
+ \bar q_L^\alpha (\Lambda_{q_L'})_\alpha^I ({\cal O}_{q_L'})_I
+\bar u_R (\Lambda_{u_R})^I ({\cal O}_{u_R})_I
+ \bar d_R (\Lambda_{d_R})^I ({\cal O}_{d_R})_I\,,
\end{equation}
$\alpha$ being an $\SU(2)_L$ index and $I$ an $\SO(6)$ index.
The corresponding spurions are given by
\bea
&& \Lambda_{q_L}^{\bold 6} = \frac{\lambda_{q_L}}{\sqrt{2}} \left( \begin{array}{cccccc}
0 & 0 & i & -1 & 0 & 0 \\
-i & 1 & 0 & 0 & 0 & 0
\end{array} \right)^T\,, \qquad 
\Lambda_{u_R}^{\bold 6} = \lambda_{u_R} \left( \begin{array}{cccccc} 
0 & 0 & 0 & 0 &  e^{i \alpha_{u_6}} \sin \theta_{u_6} & \cos \theta_{u_6}
\end{array}
\right)^T\,, \nn \\
&& \Lambda_{q_L'}^{\bold 6} = \frac{\lambda_{q_L'}}{\sqrt{2}} \left( \begin{array}{cccccc}
i & 1 & 0 & 0 & 0 & 0 \\
0 & 0 & -i & -1 & 0 & 0
\end{array} \right)^T\,, \qquad 
\Lambda_{d_R}^{\bold 6} = \lambda_{d_R} \left( \begin{array}{cccccc} 
0 & 0 & 0 & 0 &  e^{i \alpha_{d_6}} \sin \theta_{d_6} & \cos \theta_{d_6}
\end{array}
\right)^T\,.\label{eq:ferm_spurions}
\eea
Notice that the embedding of the $q_L$ doublet into the ${\bf 6}_{+2/3}$ and ${\bf 6}_{-1/3}$ representations is fixed.
On the contrary there are two independent embeddings for the right-handed quarks, which can populate both
the 5-th and the 6-th component of the $\SO(6)$ multiplet. The admixture is parametrised by the angles $\theta_{u_6},\theta_{d_6}$ and by the complex phases $\alpha_{u_6}, \alpha_{d_6}$.
If $\theta_{u_6, d_6} = \pi/2$ the couplings of the right-handed fermions do not break the $\SO(5)_H$ subgroup
(see sec.~\ref{sec:coset}) and do not give rise to a potential for the Higgs doublet. In this case it is difficult to realise
a phenomenologically viable scenario.

The operators contributing to the effective potential can be obtained
by constructing invariant combinations using the spurions in eq.~(\ref{eq:ferm_spurions}) and the Goldstone
multiplet $\Sigma$~\cite{DeSimone:2012fs,Panico:2015jxa}. We can classify them in powers of the elementary--composite mixing parameters,
$\lambda_{q_L, q_L', u_R, d_R}$.

At the quadratic level there are only four independent invariants. Namely the two combinations
\begin{eqnarray}
{\cal O}^{(2)}_{q_L} &\equiv& \Sigma^T\!\cdot\!{\Lambda_{q_L}^{\bf 6}}\cdot{\Lambda_{q_L}^{\bf 6}}^\dagger\!\cdot\!\Sigma
= \frac{1}{2} |\lambda_{q_L}|^2 s_h^2\,,\label{eq:O2qL}\\
{\cal O}^{(2)}_{u_R} &\equiv& \Sigma^T\!\cdot\!{\Lambda_{u_R}^{\bf 6}}\; {\Lambda_{u_R}^{\bf 6}}^\dagger\!\cdot\!\Sigma
= |\lambda_{u_R}|^2 \left|f_{u_6}(h, \eta)\right|^2\,,\label{eq:O2uR}
\end{eqnarray}
where
\begin{equation}
f_{u_6}(h, \eta) \equiv \cos \theta_{u_6} \sqrt{1- s_h^2 - s_\eta^2}
+ e^{i \alpha_{u_6}} \sin \theta_{u_6} s_\eta\,,
\end{equation}
and the analogous ones built from the $\Lambda_{q_L'}^{\bf 6}$ and $\Lambda_{d_R}^{\bf 6}$ spurions.
The coefficients multiplying the quadratic invariants generated at one-loop level
are of order $\frac{N_c}{16 \pi^2} m_\Psi^2$, where $m_\Psi$ denotes the mass scale of the fermionic
partners.\footnote{For simplicity we do not distinguish between the mass scale of the top partners and of the bottom
partners.}

The ${\cal O}_{q_L}^{(2)}$ and ${\cal O}_{q_L'}^{(2)}$ invariants only depend on the Higgs doublet and not on $\eta$, since
the $\Lambda_{q_L}^{\bf 6}$ and $\Lambda_{q_L'}^{\bf 6}$ spurions do not break the $\SO(2)_\eta$ symmetry.
The structure of the ${\cal O}_{u_R}^{(2)}$ and ${\cal O}_{d_R}^{(2)}$ invariants is instead more complex. For definiteness we focus on
${\cal O}_{u_R}^{(2)}$. As we mentioned
before, for $\theta_{u_6} = \pi/2$ the $\Lambda_{u_R}^{\bf 6}$ spurion does not break the $\SO(5)_H$ symmetry
and does not generate a potential for the Higgs doublet.
Let us now focus on the $P_\eta$ symmetry. Two obvious cases in which $P_\eta$ is preserved correspond to
the choices $\theta_{u_6} = 0$ and $\theta_{u_6} = \pi/2$, in which the $t_R$ is embedded in only one of the last
two components of the $\SO(6)$ multiplet.
There is however an additional case in which the ${\cal O}_{u_R}^{(2)}$ invariant is
symmetric under $P_\eta$, namely when $\alpha_{u_6} = \pm\pi/2$ (regardless of the value of $\theta_{u_6}$).
This is a consequence of a more general result which shows that if the composite sector is invariant under
${\rm O}(5)$ and CP, then the effective potential is automatically invariant under $P_\eta$. Even if the composite
dynamics is not invariant under ${\rm O}(5)$ and CP, the lowest order invariants of the potential accidentally
respect $P_\eta$, in particular, in the $({\bf 6}, {\bf 6})$ model, this is valid for all invariants up to quartic order.
A proof of these results is given in appendix~\ref{app:p_eta}.

Another special configuration is the one in which $\alpha_{u_6, d_6} = \pm\pi/2$ and $\theta_{u_6, d_6} = \pi/4$. In this case
under an $\SO(2)_\eta$ transformation the $\Lambda_{u_R}^{\bf 6}$ and $\Lambda_{d_R}^{\bf 6}$ spurions rotate by an overall phase, which
can be compensated by a corresponding phase redefinition of $t_R$ and $b_R$. The whole Lagrangian is thus invariant
under $\SO(2)_\eta$ and no potential for the $\eta$ singlet is generated (this can be easily verified explicitly for the
${\cal O}_{u_R}^{(2)}$ invariant in eq.~(\ref{eq:O2uR})).

At the quartic level, six additional independent invariants appear at one loop in the effective potential.\footnote{We neglect
invariants that are trivially obtained from the quadratic ones multiplying them by a constant factor of order $\lambda^2$.}
The ones involving the $\Lambda_{q_L}^{\bf 6}$ and $\Lambda_{u_R}^{\bf 6}$ spurions are
\begin{eqnarray}
{\cal O}^{(4)}_{q_L} &\equiv& (\Sigma^T\cdot{\Lambda_{q_L}^{\bf 6}}\cdot{\Lambda_{q_L}^{\bf 6}}^\dagger\cdot\Sigma)^2
= \frac{1}{4} |\lambda_{q_L}|^4 s_h^4\,,\label{eq:O4qL}\\
{\cal O}^{(4)}_{u_R} &=& (\Sigma^T\cdot{\Lambda_{u_R}^{\bf 6}}\; {\Lambda_{u_R}^{\bf 6}}^\dagger\!\cdot\!\Sigma)^2
= |\lambda_{u_R}|^4 \left|f_{u_6}(h, \eta)\right|^4\,,\label{eq:O4uR}\\
{\cal O}^{(4)}_{q_L u_R} &\equiv& (\Sigma^T\!\cdot\!{\Lambda_{q_L}^{\bf 6}}\cdot{\Lambda_{q_L}^{\bf 6}}^\dagger\!\cdot\!\Sigma)(\Sigma^T\!\cdot\!{\Lambda_{u_R}^{\bf 6}}\; {\Lambda_{u_R}^{\bf 6}}^\dagger\!\cdot\!\Sigma)
= \frac{1}{2} |\lambda_{q_L}|^2 |\lambda_{u_R}|^2 s_h^2 \left|f_{u_6}(h, \eta)\right|^2\,,\label{eq:O4qLuR}
\end{eqnarray}
and the remaining three are analogously built from $\Lambda_{q_L'}^{\bf 6}$ and $\Lambda_{d_R}^{\bf 6}$. With
respect to the quadratic invariants, the coefficients of the quartic ones are suppressed by a factor of order
$f^2/m_\Psi^2$ and are thus of order $\frac{N_c}{16\pi^2}f^2$.

\begin{table}
\centering
\begin{tabular}{c||c|c||c|c|c}
$({\bf 6}, {\bf 6})$ & ${\cal O}_{q_L}^{(2)}$ & ${\cal O}_{u_R}^{(2)}$ & ${\cal O}_{q_L}^{(4)}$ & $\widetilde {\cal O}_{u_R}^{(4)}$ & $\widetilde {\cal O}_{q_L u_R}^{(4)}$\\
\hline\hline
\rule[-10pt]{0pt}{2.25em}$\frac{N_c}{16 \pi^2}\times$ & $\lambda_{q_L}^2 \frac{m_\Psi^2}{f^2}$ & $-2 \lambda_{u_R}^2 \frac{m_\Psi^2}{f^2}$ & $\lambda_{q_L}^4$ & $4 \lambda_{u_R}^4$ & $\lambda_{q_L}^2 \lambda_{u_R}^2$\\
\hline\hline
\rule[-7pt]{0pt}{2em}$\mu_h^2/f^2$ & $1$ & $\cos^2 \theta_{u_6}$ & $0$ & $0$ & $0$\\
\rule[-7pt]{0pt}{2em}$\mu_\eta^2/f^2$ & $0$ & $\cos 2 \theta_{u_6}$ & $0$ & $0$ & $0$\\
\hline
\rule[-7pt]{0pt}{2em}$\lambda_h$ & $-\frac{2}{3}$ & $-\frac{2}{3} \cos^2 \theta_{u_6}$ & $1$ & $\cos^4 \theta_{u_6}$ & $-2\cos^2 \theta_{u_6}$\\
\rule[-7pt]{0pt}{2em}$\lambda_\eta$ & $0$ & $-\frac{2}{3} \cos 2 \theta_{u_6}$ & $0$ & $\cos^2 2 \theta_{u_6}$ & $0$\\
\rule[-7pt]{0pt}{2em}$\lambda_{h\eta}$ & $0$ & $0$ & $0$ & $\cos^2 \theta_{u_6} \cos 2 \theta_{u_6}$ & $-\cos 2\theta_{u_6}$
\end{tabular}
\caption{Contributions to the coefficients of the effective potential coming from the quadratic and quartic one-loop invariants
from fermions in the fundamental representation of $\SO(6)$. The results correspond to the $P_\eta$-symmetric case, $\alpha_{u_6} = \pm\pi/2$. We list only the invariants coming from the
$\lambda_{q_L}$ and $\lambda_{u_R}$ mixings, the ones from the bottom sector are analogous.
The second line reports the estimate of the size of the coefficients multiplying each invariant.\label{tab:invariants_6}}
\end{table}

The contributions of the quadratic and quartic invariants to the coefficients of the effective potential
defined in eq.~(\ref{eq:V_elem}) are listed in table~\ref{tab:invariants_6}. Notice that, in order to simplify the
results, we redefined the ${\cal O}^{(4)}_{u_R}$ and ${\cal O}^{(4)}_{q_L u_R}$ by a shift, which however
does not affect the power counting estimates:\footnote{This redefinition is fully equivalent to a shift of the coefficients
of the quadratic invariants by a quantity of order $\lambda^4$.}
\begin{eqnarray}
\widetilde {\cal O}^{(4)}_{u_R} &\equiv& {\cal O}^{(4)}_{u_R} - 2 \cos^2 \theta_{u_6} |\lambda_{u_R}|^2 {\cal O}^{(2)}_{u_R}\,,
\label{eq:O4R6_shifted}\\
\rule{0pt}{1.15em}\widetilde {\cal O}^{(4)}_{q_L u_R} &\equiv& {\cal O}^{(4)}_{q_L u_R} - \cos^2 \theta_{u_6} |\lambda_{q_L}|^2 {\cal O}^{(2)}_{q_L}\label{eq:O4LR6_shifted}\,.
\end{eqnarray}
In table~\ref{tab:invariants_6} we only considered the,
phenomenologically more appealing, $P_\eta$-symmetric case obtained for $\alpha_{u_6} = \pm\pi/2$.
The reason for this choice is the following. If $P_\eta$ is broken, a tadpole term for $\eta$ is generated
and the singlet always gets a VEV.  On top of that, by inspection of the interactions with the fermions, the breaking
of $P_\eta$ is accompanied by an explicit breaking of CP, which can generate dangerously large effects in flavour
observables.
For these reasons, we focus on the scenario in which, at zero temperature, only the Higgs doublet gets a VEV
while $\eta$ does not and we consider the case $\alpha_{u_6} = \pm \pi/2$.

The size of the $\lambda$ parameters can be estimated by connecting them with the mass of the top and bottom quarks.
The operator that describes the mass of the top and its interactions with the Goldstones is given by
\begin{eqnarray}
{\cal O}_t &=& - c_t \frac{f^2}{m_\Psi} \bar q_L^\alpha (\Lambda^{\bf 6}_{q_L})_\alpha\!\cdot\!\Sigma\;\Sigma^T\!\cdot\!{\Lambda^{\bf 6}_{u_R}}^\dagger t_R
+ {\rm h.c.}\nn\\
&=& \frac{c_t}{\sqrt{2}} \frac{f^2}{m_\Psi} \lambda_{q_L} \lambda_{u_R}^* s_h\, f^*_{u_6}(h, \eta)\,\bar t_L t_R + {\rm h.c.}\,,
\end{eqnarray}
with $c_t$ an order-one coefficient. Expanding in the Higgs and singlet fields we find
\begin{equation}
{\cal O}_t = \frac{c_t \lambda_{q_L} \lambda_{u_R}^* f}{\sqrt{2} m_\Psi} \left(\cos \theta_{u_6}\, h + \frac{e^{-i \alpha_{u_6}}}{f}
\sin \theta_{u_6}\, \eta\, h\right) \bar t_L t_R + {\rm h.c.}\,,
\end{equation}
so that we find the relation of the $\lambda$ coefficients with the top Yukawa $y_t$
\begin{equation}\label{eq:top_Yukawa}
\lambda_{q_L} \lambda_{u_R}^* \frac{f}{m_\Psi} \cos \theta_{u_6} \simeq y_t\,.
\end{equation}
Notice that the ${\cal O}_t$ operator gives also rise to a $\eta\,h\,\bar t_L\,t_R$ interaction, which, as we will
discuss later on, can be important as a source of CP violation.

With similar steps one finds the following estimate for the bottom Yukawa $y_b$,
\begin{equation}\label{eq:bottom_Yukawa}
\lambda_{q_L'} \lambda_{d_R}^* \frac{f}{m_\Psi} \cos \theta_{d_6} \simeq y_b\,.
\end{equation}

\bigskip

Let's now discuss the features of the effective potential. The leading contributions are typically the ones coming from
the top quark sector, so we start by focusing on their properties.
The structure of the Higgs doublet potential is completely analogous to the one obtained in minimal
$\SO(5)/\SO(4)$ models~\cite{Matsedonskyi:2012ym}. Depending on the mass scale of the top partners, we can identify two different regimes:
a heavy-partner regime in which $m_\Psi/f \gtrsim 3$ and a light-partner regime in which $m_\Psi/f \sim 1$.

In the heavy-partner case the quadratic invariants dominate the potential and a tuning between ${\cal O}^{(2)}_{q_L}$ and
${\cal O}^{(2)}_{u_R}$ is required to obtain the correct Higgs mass, namely $\lambda_{q_L} \simeq \sqrt{2} \cos \theta_{u_6} \lambda_{u_R}$.\footnote{Notice that even if the overall top-partner scale is heavy, at least one accidentally-light top partner
is still needed to obtain a viable Higgs potential~\cite{Matsedonskyi:2012ym,Redi:2012ha,Marzocca:2012zn,Pomarol:2012qf}.}
The estimate of the top Yukawa in eq.~(\ref{eq:top_Yukawa}) implies that
$\lambda_{u_R} \sim \lambda_{q_L} \simeq y_t g_\rho$, where $g_\rho \equiv m_\Psi/f$ can be identified with the
typical coupling of the composite dynamics. The quartic invariants are thus suppressed with respect to the
quadratic ones by a factor $\lambda^2 f^2/m_\Psi^2 \sim 1/g_\rho$.
The two-step transition conditions in eq.~(\ref{eq:conditions_eta}) are difficult to realise in this limit.
As can be seen from table~\ref{tab:invariants_6}, the portal contribution to the mass $\lambda_{h\eta} v^2$
is suppressed with respect to the $\mu_\eta^2$ term by a factor $\lambda^2 v^2/m_\Psi^2 \sim 1/g_\rho v^2/f^2$.
Therefore, if $\mu_\eta^2$ is negative, it is difficult to avoid a VEV for the singlet at zero temperature.

A possibility to circumvent this problem is to advocate a sizeable contribution to the potential from the bottom sector.
This can be obtained if both the top and the bottom quarks have a large compositeness for their right-handed components,
namely $\lambda_{u_R} \sim \lambda_{d_R}$. The mass of the bottom quark is then reproduced by assuming that
$\lambda_{q_L'}$ is small. This scenario, however, could lead to difficulties in realising the CKM hierarchy structure.

In the light-partner case, all the invariants are of the same order, therefore it is much easier to obtain the correct Higgs mass
and satisfy the two-step transition conditions. The price to pay is the fact that all top partners are now typically light and
higher values of the compositeness scale $f$ are needed to escape LHC direct-search constraints. A larger amount of
tuning, $\xi = v^2/f^2 \lesssim \textit{few}\,\%$, is therefore needed to obtain the correct Higgs VEV.

Due to the form of the invariants, sharp upper bounds on the portal coupling $\lambda_{h\eta}$ and on the singlet mass
in the EWSB vacuum can be found, namely
\begin{equation}\label{eq:constr_6-6}
\lambda_{h\eta} < \lambda_h\,, \qquad \quad m_\eta < m_h/\sqrt{2}\,.
\end{equation}
To prove the first inequality one needs to use the fact that the coefficients of the ${\cal O}^{(4)}_{q_L}$
and $\widetilde {\cal O}^{(4)}_{u_R}$ invariants are always positive, while the coefficient of $\widetilde {\cal O}^{(4)}_{q_L u_R}$
is negative. This result can be obtained by studying the explicit form of the effective potential as
done in ref.~\cite{Frigerio:2012uc}. The $\widetilde {\cal O}^{(4)}_{u_R}$ invariant thus gives contributions
$\Delta \lambda_{h\eta} \leq \Delta \lambda_h$,\footnote{To ensure that $\lambda_{h\eta}$ is positive
one needs $\sin 2\theta_{u_6} > 0$, which implies $\cos 2 \theta_{u_6} \leq \cos^2\theta_{u_6}$.} while $\widetilde {\cal O}^{(4)}_{q_L u_R}$ gives
$\Delta \lambda_{h\eta} \leq 1/2 \Delta \lambda_h$. The ${\cal O}^{(4)}_{q_L}$ invariant provides only a positive
contribution to $\lambda_h$. Finally the sum of the quadratic invariants
give $\Delta \lambda_h = -2/3 \Delta \mu_h^2/f^2 > 0$, since the Higgs mass term must be negative.

The second inequality in eq.~(\ref{eq:constr_6-6}) can be derived by noticing that in the EWSB vacuum
\begin{equation}
m_\eta^2 = \mu_\eta^2 + \lambda_{h\eta} v^2 < \lambda_{h\eta} v^2 < \lambda_h v^2 = m_h^2/2\,.
\end{equation}
The bound on the singlet mass is particularly dangerous since it implies that the singlet is always quite light.
In particular in a sizeable part of the parameter space $m_\eta < m_h/2$ and the Higgs is allowed to decay into a pair
of singlets. These configurations are excluded experimentally since they would give rise to a too large
invisible width for the Higgs unless the portal coupling is negligibly small.

\subsubsection{Fermions in the ($\bold{15}$) representation}
\label{sec:(15,6)-model}

As we saw, in the minimal set-up with partners in the fundamental representation of $\SO(6)$ it is quite hard to obtain
a two-step phase transition. Moreover the size of the portal interaction is severely constrained. We consider now a
next-to-minimal case in which some of the partners belong to the $\bf 15$ representation.

The $\bold{15}$ representation can be obtained as the antisymmetric product of two $\bold 6$. As can be seen from
eq.~(\ref{eq:reps}), the $\bold{15}$ provides two embeddings for both the left-handed and the right-handed top quark
components. The corresponding spurions are
\bea
\left(\Lambda_{q_L}^{\bold{15}}\right)_\alpha &=& \frac{\lambda_{q_L}}{2}\left(   \begin{array}{ccc}
0_{4 \times 4}  & e^{i \alpha_{q_{15}}} \sin \theta_{q_{15}} \, \vec{v}_\alpha & \cos \theta_{q_{15}} \, \vec{v}_\alpha \\
- e^{i \alpha_{q_{15}}} \sin \theta_{q_{15}} \, \vec{v}_\alpha^T & 0 & 0 \\
- \cos \theta_{q_{15}} \, \vec{v}_\alpha^T & 0 & 0
\end{array} 
\right)
\,, \nn \\ 
 \Lambda_{u_R}^{\bold{15}} &=& \frac{\lambda_{u_R}}{\sqrt{2}} \left( \begin{array}{ccc}
\cos \theta_{u_{15}} \frac{i \sigma_2}{\sqrt{2}} & 0_{2 \times 2} & 0_{2 \times 2} \\
0_{2 \times 2} & \cos \theta_{u_{15}} \frac{i \sigma_2}{\sqrt{2}} &  0_{2 \times 2} \\
0_{2 \times 2} & 0_{2 \times 2} & e^{i \alpha_{u_{15}}} \sin \theta_{u_{15}}  \, i \sigma_2
\end{array} \right)\,,
\eea
where $\alpha = 1,2$ is an $\SU(2)_L$ doublet index and
\bea
\vec{v}_1 = (0,0,i,-1)^T \,, \qquad \vec{v}_2 = (-i,1,0,0)^T\,.
\eea

The easiest possibility to achieve a two-step first order phase transition is to consider the
mixed embedding $(q_L, u_R) \sim (\bold{15}, \bold 6)$. Other choices lead to more contrived scenarios.
In fact, the $(\bold{15}, \bold 1)$ embedding does not allow for a mass term for the top quark, the $(\bold 6, \bold {15})$ does not generate a potential for the singlet , while in the $(\bold{15}, \bold{15})$ case is difficult to avoid
$\lambda_\eta \simeq 0$.

The quadratic invariants in the $(\bold{15}, \bold 6)$ model are given by
\begin{eqnarray}
{\cal O}^{(2)}_{q_L} &\equiv& \Sigma^T\!\cdot\left({\Lambda_{q_L}^{\bf 15}}\right)_\alpha^\dagger\cdot\left(\Lambda_{q_L}^{\bf 15}\right)^\alpha\cdot\!\Sigma
= \frac{1}{2} |\lambda_{q_L}|^2 \left(f^u_{q_{15}}(h, \eta) + f^d_{q_{15}}(h, \eta)\right)\,,\label{eq:O2qL15} \\
{\cal O}^{(2)}_{u_R} &\equiv& \Sigma^T\!\cdot\!{\Lambda_{u_R}^{\bf 6}}\; {\Lambda_{u_R}^{\bf 6}}^\dagger\!\cdot\!\Sigma
= |\lambda_{u_R}|^2 \left|f_{u_6}(h, \eta)\right|^2\,,
\end{eqnarray}
where
\begin{eqnarray}
f^u_{q_{15}}(h, \eta) &\equiv& \frac{1}{2}\left(1 - s_h^2 - 2 s_\eta^2\right)
\cos 2 \theta_{q_{15}} + s_\eta \sqrt{1- s_h^2 - s_\eta^2} \cos \alpha_{q_{15}} \sin 2 \theta_{q_{15}} \,,\\
f^d_{q_{15}}(h, \eta) &\equiv& \frac{1}{2} s_h^2 + f^u_{q_{15}}(h, \eta)\,.
\end{eqnarray}
The quartic invariants are given by
\begin{eqnarray}
{\cal O}^{(4)}_{q_L} &\equiv& \left(\Sigma^T\!\cdot\left({\Lambda_{q_L}^{\bf 15}}\right)_\alpha^\dagger\cdot\left(\Lambda_{q_L}^{\bf 15}\right)^\beta\cdot\!\Sigma\right) \left(\Sigma^T\!\cdot\left({\Lambda_{q_L}^{\bf 15}}\right)_\beta^\dagger\cdot\left(\Lambda_{q_L}^{\bf 15}\right)^\alpha\cdot\!\Sigma\right) = \nn\\
&=& \frac{1}{4} |\lambda_{q_L}|^4 \left(\left(f^u_{q_{15}}(h, \eta)\right)^2 + \left(f^d_{q_{15}}(h, \eta)\right)^2\right)\label{eq:O4qL15}\\
{\cal O}^{(4)}_{u_R} &=& (\Sigma^T\cdot{\Lambda_{u_R}^{\bf 6}}\; {\Lambda_{u_R}^{\bf 6}}^\dagger\!\cdot\!\Sigma)^2
= |\lambda_{u_R}|^4 \left|f_{u_6}(h, \eta)\right|^4\,,\\
{\cal O}^{(4)}_{q_L u_R} &\equiv& \left(\Sigma^T\!\cdot\left({\Lambda_{q_L}^{\bf 15}}\right)_\alpha^\dagger\cdot
\Lambda_{u_R}^{\bf 6}\right)
({\Lambda_{u_R}^{\bf 6}}^\dagger\!\cdot\!\left(\Lambda_{q_L}^{\bf 15}\right)^\alpha\!\cdot\!\Sigma)\nn\\
&=& \frac{1}{4} |\lambda_{q_L}|^2 |\lambda_{u_R}|^2 \left|\cos \theta_{q_{15}} \cos \theta_{u_6}
+ e^{i (\alpha_{q_{15}} - \alpha_{u_6})} \sin \theta_{q_{15}} \sin \theta_{u_6}\right|^2 s_h^2\,.\label{eq:O4qLuR15}
\end{eqnarray}
Notice that the ${\cal O}^{(2)}_{u_R}$ and ${\cal O}^{(4)}_{u_R}$ invariants obviously coincide with the ones we found in the $({\bf 6}, {\bf 6})$ model.

Similarly to what happens in the $({\bf 6}, {\bf 6})$ model, the choice $\alpha_{q_{15}} = \pm \pi/2$ and
$\alpha_{u_6} = \pm \pi/2$ guarantees that the potential respects the $P_\eta$ invariance up to quartic order.
The coefficients of the effective potential for the $P_\eta$-symmetric case are listed in table~\ref{tab:invariants_15}.
Similarly to what we did in table~\ref{tab:invariants_6}, we shifted the quartic invariants to simplify the results.
The $\widetilde {\cal O}^{(4)}_{u_R}$ invariant is defined in eq.~(\ref{eq:O4R6_shifted}), while $\widetilde {\cal O}^{(4)}_{q_L}$
is given by
\begin{equation}
\widetilde {\cal O}^{(4)}_{q_L} = {\cal O}^{(4)}_{q_L} - \cos^2 \theta_{q_{15}} |\lambda_{q_L}|^2 {\cal O}^{(2)}_{q_L}\,.
\end{equation}

\begin{table}
\centering
\footnotesize
\begin{tabular}{@{\,}c@{\;}||@{\;}c@{\;}|@{\;}c@{\;}||@{\;}c@{\;}|@{\;}c@{\;}|@{\;}c@{\,}}
$({\bf 15}, {\bf 6})$ & ${\cal O}_{q_L}^{(2)}$ & ${\cal O}_{u_R}^{(2)}$ & $\widetilde {\cal O}_{q_L}^{(4)}$ & $\widetilde {\cal O}_{u_R}^{(4)}$ & ${\cal O}_{q_L u_R}^{(4)}$\\
\hline\hline
\rule[-8pt]{0pt}{2.25em}$\frac{N_c}{16 \pi^2}\times$ & $-\lambda_{q_L}^2 \frac{m_\Psi^2}{f^2}$ & $-2 \lambda_{u_R}^2 \frac{m_\Psi^2}{f^2}$ & $\frac{1}{2}\lambda_{q_L}^4$ & $4 \lambda_{u_R}^4$ & $\frac{1}{2}\lambda_{q_L}^2 \lambda_{u_R}^2$\\
\hline\hline
\rule[-7pt]{0pt}{2em}$\mu_h^2/f^2$ & $\frac{1}{2} + \cos 2 \theta_{q_{15}}$ & $\cos^2 \theta_{u_6}$ & $0$ & $0$ & $\cos^2(\theta_{q_{15}} - \theta_{u_6})$\\
\rule[-7pt]{0pt}{2em}$\mu_\eta^2/f^2$ & $2 \cos 2 \theta_{q_{15}}$ & $\cos 2 \theta_{u_6}$ & $0$ & $0$ & $0$ \\
\hline
\rule[-7pt]{0pt}{2em}$\lambda_h$ & $-\frac{1}{3} - \frac{2}{3} \cos 2 \theta_{q_{15}}$ & $-\frac{2}{3} \cos^2 \theta_{u_6}$ &  $\frac{1}{4}\left(1+\left(1 + 2\cos 2\theta_{q_{15}}\right)^2\right)$ & $\cos^4 \theta_{u_6}$ & $-\frac{2}{3}\cos^2(\theta_{q_{15}} - \theta_{u_6})$\\
\rule[-7pt]{0pt}{2em}$\lambda_\eta$ & $-\frac{4}{3} \cos 2 \theta_{q_{15}}$ & $-\frac{2}{3} \cos 2 \theta_{u_6}$ & $4 \cos^2 2 \theta_{q_{15}}$ & $\cos^2 2 \theta_{u_6}$ & $0$\\
\rule[-7pt]{0pt}{2em}$\lambda_{h\eta}$ & $0$ & $0$ & $\cos 2 \theta_{q_{15}} (1 + 2 \cos 2 \theta_{q_{15}})$ & $\cos^2 \theta_{u_6} \cos 2 \theta_{u_6}$ & $0$
\end{tabular}
\caption{Fermion contributions to the coefficients of the effective potential coming from the quadratic and quartic one-loop invariants in the $({\bf 15}, {\bf 6})$ model. The results correspond to the $P_\eta$-symmetric case $\alpha_{q_{15}} = \alpha_{u_6} =\pm\pi/2$. In the case $\alpha_{q_{15}} = -\alpha_{u_6} = \pm\pi/2$ the results can be obtained by reversing the sign of $\theta_{u_6}$.
\label{tab:invariants_15}}
\end{table}

The operator that describes the mass of the top and its interactions with the Goldstones is given by
\begin{eqnarray}
{\cal O}_t &=& - c_t \frac{f^2}{m_\Psi} \bar q_L^\alpha\; {\Lambda^{\bf 6}_{u_R}}^\dagger\!\cdot\!(\Lambda^{\bf 15}_{q_L})_\alpha\!\cdot\!\Sigma\; t_R
+ {\rm h.c.}\label{eq:top_Yukawa_15} \nn \\
&=& \frac{c_t \lambda_{q_L} \lambda_{u_R}^* f^2}{2 m_\Psi^2} \left(\cos \theta_{q_{15}} \cos \theta_{u_6} + e^{i(\alpha_{q_{15}} - \alpha_{u_6})} \sin \theta_{q_{15}} \sin \theta_{u_6}\right) s_h\, \bar t_L t_R + {\rm h.c.}\,,
\end{eqnarray}
with $c_t$ an order-one coefficient. The top Yukawa coupling is thus given by
\begin{equation}
y_t = \frac{c_t \lambda_{q_L} \lambda_{u_R}^* f}{\sqrt m_\Psi^2} \left|\cos \theta_{q_{15}} \cos \theta_{u_6} + e^{i(\alpha_{q_{15}} - \alpha_{u_6})} \sin \theta_{q_{15}} \sin \theta_{u_6}\right|\,.
\end{equation}
In the $P_\eta$-symmetric case this result simplifies as
\begin{equation}
y_t = \frac{c_t \lambda_{q_L} \lambda_{u_R}^* f}{\sqrt{2} m_\Psi^2} \cos (\theta_{q_{15}} - \theta_{u_6})\,,
\end{equation}
where we chose $\alpha_{q_{15}} = \alpha_{u_6} = \pm \pi/2$ (the case $\alpha_{q_{15}} = -\alpha_{u_6} = \pm \pi/2$
can be obtained by reversing the sign of $\theta_{u_6}$ in the formula).
It is interesting to notice that the operator in eq.~(\ref{eq:top_Yukawa_15}) only contains the Higgs field and not the singlet.
Therefore, at the leading order in the $\lambda$ expansion, no interaction of the form $\eta\,h\,\bar t_L\, t_R$ is present
in the $({\bf 15}, {\bf 6})$ model.

\bigskip

Let's now study the properties of the effective potential and the conditions for a two-step EWPhT
in the $({\bf 15}, {\bf 6})$ model. As for the $({\bf 6}, {\bf 6})$ set-up, we can distinguish two regimes, the heavy-partner
limit and the light-partner one.

In the heavy-partner case, the quadratic invariants dominate the effective potential. With respect the $({\bf 6}, {\bf 6})$ model,
however, there is a substantial difference, namely the fact that the singlet mass term receives contributions from both leading
invariants and not just one. This means that, at the price of some additional tuning, the Higgs mass and the $\eta$ mass
term can be simultaneously cancelled. For this to happen we need a correlation between the left and right mixing parameters $\lambda_{q_L} \simeq \lambda_{u_R}$ and between the embedding angles $\theta_{q_{15}}$ and $\theta_{u_6}$. Once the Higgs mass is tuned, a cancellation in the $\mu_\eta$ term can be obtained if $\sin \theta_{q_{15}} \simeq 1/3 (3/2 + \sin^2 \theta_{u_6})$,
which can be realized only if $\theta_{q_{15}}$ is the range $0.5 \lesssim \theta_{q_{15}} \lesssim 1$. If both cancellations are
present, it is then easy to satisfy the two-step conditions in eq.~(\ref{eq:conditions_eta}), through a positive $\lambda_{h\eta}$
term. In this set-up, however, the portal interaction can not surpass the Higgs quartic coupling, $\lambda_{h\eta} < \lambda_h$.
Indeed, taking into account
the restricted range of $\theta_{q_{15}}$ values, one finds that for both the $\widetilde {\cal O}^{(4)}_{u_R}$ and $\widetilde {\cal O}^{(4)}_{q_L}$
invariants $\Delta \lambda_{h\eta} < \Delta \lambda_h$. As a consequence one also gets $m_\eta < m_h/\sqrt{2}$.

In the light-partner case, additional contributions to the Higgs mass term can come from the ${\cal O}^{(4)}_{q_L u_R}$
operator, moreover the quartic operators that contribute to the portal interaction are only mildly suppressed with respect to
the quadratic contributions. For these reasons a viable Higgs mass together with a two-step EWPhT can be
obtained for a larger range of values of the embedding angles $\theta_{q_{15}}$ and $\theta_{u_6}$. Also in this case
a maximal value for the portal interaction is present, namely $\lambda_{h\eta} < 2 \lambda_h$, which implies that
the singlet is always lighter than the Higgs $m_\eta < m_h$. The maximal value for the portal interaction is obtained
when the dominant contribution to $\lambda_{h\eta}$ comes from the $\widetilde {\cal O}^{(4)}_{u_R}$ invariant and $\theta_{q_{15}} \simeq \pi/2$, in which case $\Delta \lambda_{h\eta} = 2 \Delta \lambda_h$.

Summarising, differently from the previous case with fermions in the fundamental of $\SO(6)$,
in the $(\bold{15}, \bold 6)$ model viable configurations with a two-step EWPhT can be realised at the price
of some tuning. The leading contribution to the potential coming from the top sector can be enough to obtain
a sufficiently large value for the portal coupling, so that sizeable contributions from the bottom (or the gauge) sectors
are not strictly necessary.

\subsubsection{Fermions in the ($\bold{20}'$) representation}
\label{sec:rep_20}

The last case we consider is the one with top partners in the ${\bf 20}'$ representation of $\SO(6)$.
This representation can be constructed as the symmetric and traceless component of the product of two $\bold 6$.
The spurions that correspond to the embedding of the left-handed and the right-handed top components are given by
\bea
 (\Lambda_{q_L}^{\bold{20}})_\alpha &=& \frac{1}{2} \lambda_{q_L}\left(   \begin{array}{ccc}
0_{4 \times 4}  & e^{i \alpha_{q_{20}}} \sin \theta_{q_{20}} \, \vec{v}_\alpha & \cos \theta_{q_{20}} \, \vec{v}_\alpha \\
 e^{i \alpha_{q_{20}}} \sin \theta_{q_{20}} \, \vec{v}_\alpha^T & 0 & 0 \\
 \cos \theta_{q_{20}} \, \vec{v}_\alpha^T & 0 & 0
\end{array} 
\right)
\,, \quad \nn \\
\Lambda_{u_R}^{\bold{20}} &=& 
\lambda_{u_R} \frac{- e^{i \alpha_{u_{20}}} \cos \theta_{u_{20}}}{2\sqrt{3}} \textrm{diag}(1,1,1,1,-2,-2) \nn \\ 
&&+\;\lambda_{u_R} \frac{\sin \theta_{u_{20}}}{\sqrt{2}}    \left(   \begin{array}{ccc}
0_{4 \times 4} & 0_{4\times 1} & 0_{4 \times 1} \\
0_{1\times 4} & e^{i \beta_{u_{20}}} \sin \phi_{u_{20}}  & \cos \phi_{u_{20}} \\
0_{1\times 4} & \cos \phi_{u_{20}}  & - e^{i \beta_{u_{20}}}  \sin \phi_{u_{20}} \\
\end{array} 
\right)\,.
\eea
where $\alpha = 1,2$ is an $\SU(2)_L$ doublet index.
Notice that there are two different embeddings for the $q_L$, parametrised by the angle $\theta_{q_{20}}$ and the complex phase $\alpha_{q_{20}}$, and three independent ones for the $u_R$, the latter described by the two angles $\theta_{u_{20}}, \phi_{u_{20}}$ and the two phases $\alpha_{u_{20}}, \beta_{u_{20}}$.

A few viable scenarios can be realised with top partners in the representation $\bold{20}'$.
One of the simplest is obtained by embedding only the right-handed top in the ${\bf 20}'$,
while assuming that the left-handed partners
transform in the fundamental representation, namely $(q_L, u_R) \sim (\bold 6, \bold{20}')$. In this case,
three independent quadratic invariants can be built
\begin{eqnarray}
{\cal O}^{(2)}_{q_L} &\equiv& \Sigma^T\!\cdot\!{\Lambda_{q_L}^{\bf 6}}\cdot{\Lambda_{q_L}^{\bf 6}}^\dagger\!\cdot\!\Sigma\,,\\
{\cal O}^{(2, 1)}_{u_R} &\equiv& \Sigma^T\!\cdot\!{\Lambda_{u_R}^{\bf 20}}^\dagger\cdot\Lambda_{u_R}^{\bf 20}\!\cdot\!\Sigma\,,\label{eq:O21uR20}\\
{\cal O}^{(2, 2)}_{u_R} &\equiv& \left|\Sigma^T\!\cdot\Lambda_{u_R}^{\bf 20}\!\cdot\!\Sigma\right|^2\,.\label{eq:O22uR20}
\end{eqnarray}
Moreover $8$ invariants at the quartic level are also present, namely one built from the left-handed spurion, three
from the right-handed spurion and four for the mixed left-right combination. The explicit form of the invariants is
reported in appendix~\ref{app:fermions_20}.

The top mass and the interactions with the Goldstones come from two independent operators,
\begin{eqnarray}
{\cal O}_t^{(1)} &=& c_t^{(1)} \frac{f^2}{m_\Psi^2} \overline q_L^\alpha \Sigma^\dagger \!\cdot\! {\Lambda^{\bf 20}_{u_R}}^\dagger \!\cdot\! (\Lambda^{\bf 6}_{q_L})_\alpha\; t_R + {\rm h.c.} \nn \\
&=& c_t^{(1)} \frac{f}{m_\Psi^2} \frac{\lambda_{q_L} \lambda_{u_R}^*}{2 \sqrt{6}} \cos \theta_{u_{20}} h\, \bar t_L t_R + {\rm h.c.} + \cdots\,,
\end{eqnarray}
and
\begin{eqnarray}
{\cal O}_t^{(2)} &=& -c_t^{(2)} \frac{f^2}{m_\Psi^2} \overline q_L^\alpha  (\Lambda^{\bf 6}_{q_L})_\alpha\!\cdot\!\Sigma\;\Sigma^T\!\cdot\!{\Lambda^{\bf 20}_{u_R}}^\dagger\!\cdot\!\Sigma\; t_R
+ {\rm h.c.} \nn \\
&=& c_t^{(2)} \frac{f \lambda_{q_L} \lambda_{u_R}^*}{m_\Psi^2}
\bigg[\left(\frac{e^{i \alpha_{u_{20}}}}{\sqrt{6}} \cos \theta_{u_{20}} \!-\! e^{i \beta_{u_{20}}} \sin \theta_{u_{20}} \sin \phi_{u_{20}}
\right) h\,\bar t_L t_R   \nn \\
&& \hspace{6em}+
\frac{1}{f}\sin \theta_{u_{20}} \cos \phi_{u_{20}}\, \eta\, h\,\bar t_L t_R \bigg] + {\rm h.c.} + \cdots\,,
\end{eqnarray}
where $c_t^{(1)}$ and $c_t^{(2)}$ are order-one coefficients. As in the other models, the top Yukawa can be estimated as
\begin{equation}
y_t \sim \frac{\lambda_{q_L} \lambda_{u_R}^* f}{m_{\Psi}}\,.
\end{equation}
Notice that the ${\cal O}_t^{(2)}$ operator gives also rise to an interaction of the form $\eta\,h\,\bar t_L t_R$.

\bigskip

We can now discuss the properties of the effective potential, whose explicit form is reported in
appendix~\ref{app:fermions_20} (table~\ref{tab:invariants_20}). With respect to the models we considered in the previous
subsections, a larger set of invariants is present in the $({\bf 6}, {\bf 20}')$ set-up. The three invariants available at quadratic
order are already enough to allow for a realistic Higgs mass and a two-step EWPhT.
The quartic terms, moreover, can provide additional freedom to enlarge the viable region of the parameter space.
Since the portal coupling can be fully disentangled from the other terms in the effective potential, its size is basically
unconstrained. This also implies that the singlet can easily be heavier than the Higgs.

\section{Parameter space for EWPhT}
\label{sec:paramspace}

\begin{figure}
\centering
\includegraphics[scale=0.6]{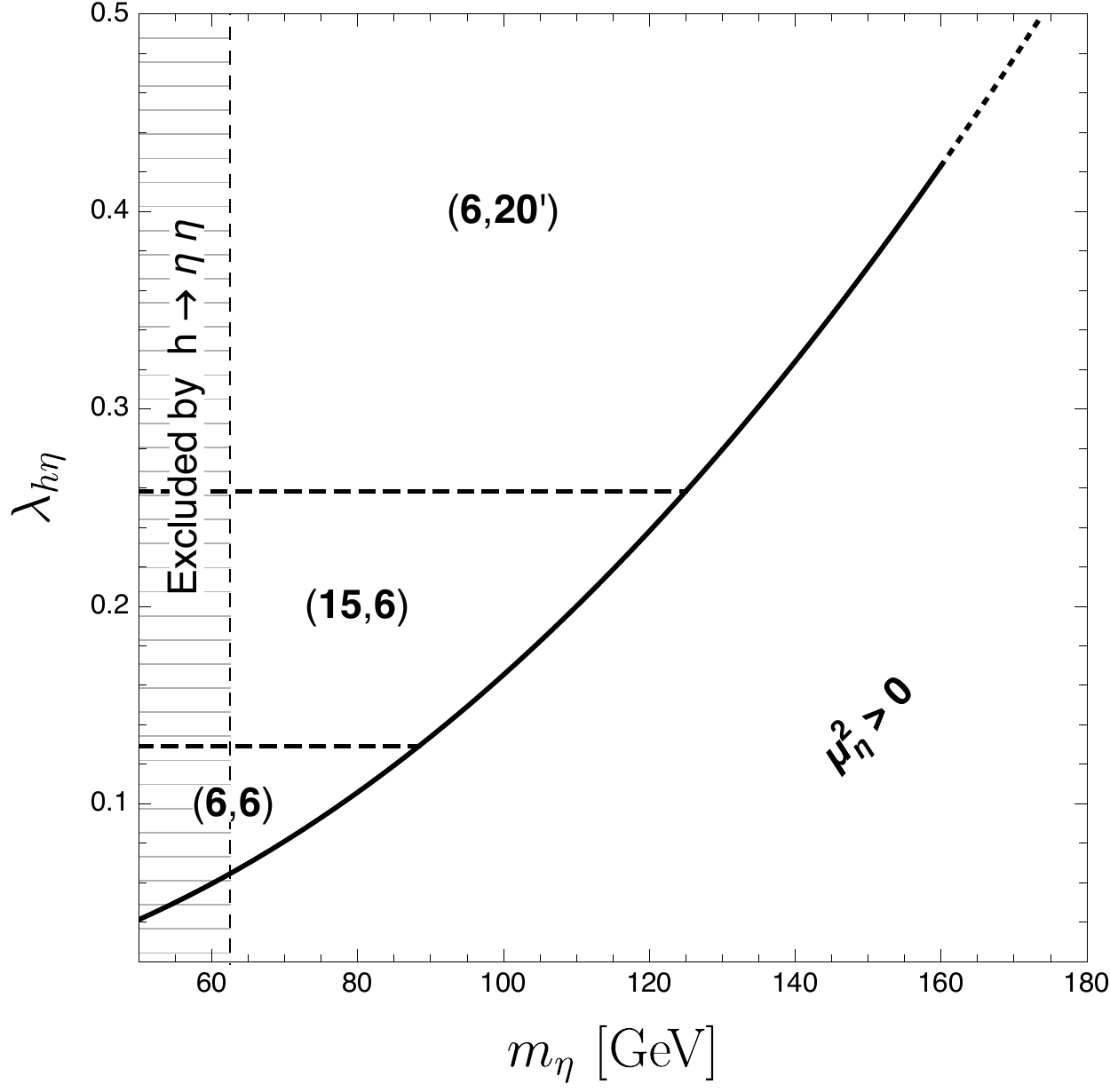}
\caption{Comparison of the parameter space covered by the different composite Higgs models discussed in the text.} 
\label{fig:reps}
\end{figure} 

We now focus our attention on the properties of the EWPhT.
The parameter space in the $(m_\eta, \lambda_{h\eta})$ plane covered by the different models discussed above
is shown in fig.~\ref{fig:reps}. 
The region on the right of the black solid line corresponds to configurations for which $\mu_\eta^2 > 0$ and
it is not interesting from the perspective of the EWPhT. In those points the singlet does not get a VEV at finite temperatures and
has a very limited impact on the vacuum structure, so that no barrier is generated between the symmetric and the EW vacua.
In this region a first order phase transition can be achieved only for very large values of the portal coupling
$\lambda_{h \eta}$ via one-loop induced effects along the Higgs direction. Since for models of composite Higgses
it is very unlikely that such large quartic couplings can be generated by the underlying strong dynamics,
we will focus our discussion only on the $\mu_\eta^2 < 0$ region in which a two-step phase transition
can be realised with a tree-level barrier.

For small singlet masses, namely $m_\eta < m_h/2$ (the dashed region in fig.~\ref{fig:reps}), the Higgs decay channel
$h \to \eta \eta$ is kinematically open and strongly affects the total decay width of the SM-like Higgs.
The invisible partial decay width of the Higgs for $m_h > 2 m_\eta$ is given by~\cite{Frigerio:2012uc}
\bea
\Gamma_{h \to \eta \eta} = \frac{v^2}{8 \pi m_h} \sqrt{1 - 4 \frac{m_\eta^2}{m_h^2}} \left( \lambda_{h \eta} \sqrt{1- \xi} - \frac{m_h^2}{2 v^2} \frac{\xi}{\sqrt{1 - \xi}} \right)^2
\eea
where $\xi = v^2/f^2$ and the second term in the expression above arises from the non-linearities of the NGB kinetic Lagrangian. 
Even though a cancellation between the two terms can be advocated in order to tame the phenomenologically
unacceptable contributions to the Higgs width, for $\xi \lesssim 0.1$ this can be achieved only for
very small values of the portal interaction, $\lambda_{h\eta} \lesssim 0.01$. Therefore these configurations do not lead
to interesting phase transition physics.
As such we will focus only on the parameter space with $m_\eta > m_h/2$.

The horizontal dashed lines in fig.~\ref{fig:reps} summarise some of the main results obtained in the previous sections.
In particular, the maximal values allowed for the quartic coupling $\lambda_{h\eta}$ in the $(\bold 6, \bold 6)$ and $(\bold{15}, \bold 6)$ models are, respectively, $\lambda_h/2$ and $2 \lambda_h$, which correspond to $m_\eta \leq m_h/\sqrt{2}$ and $m_{\eta} \leq m_h$, under the hypothesis $\mu_\eta^2 < 0$. On the other hand, no strong bound on the size of the portal coupling and of the singlet mass
is present in the $({\bf 6}, {\bf 20}')$ model.

Having defined the available parameter space for each of the three scenarios considered here, we can construct the corresponding phase transition diagrams.
In particular, one can identify the regions in which the EWPhT is a first order one and the related strength. 
For this purpose, a crucial quantity is the critical temperature $T_c$ introduced in sec.~\ref{sec:ewpht}, namely the temperature at which the two minima of the potential are degenerate.
Actually, the phase transition effectively starts at the so-called nucleation temperature $T_n$ which is lower than $T_c$ and is determined by requiring that the nucleation probability of bubbles of true vacuum per Hubble volume is of order one.
This roughly corresponds to the condition
$S_3/T_n \simeq 140$ with $S_3$ being the Euclidean action of the critical bubble. While $T_c$ can be directly calculated from the effective potential, the computation of $T_n$ requires to solve the two-field bounce equation.

The bounce configuration at finite-temperature is obtained from the minimisation of the $\textrm{O}(3)$-symmetric three-dimensional action. It corresponds to the solution of the equation
\bea
\label{eq:bounce}
\frac{d^2 \phi}{d r^2} + \frac{2}{r} \frac{d \phi}{dr} = \nabla V(\phi, T)
\eea
where $\phi = (h,\eta) $, with the boundary conditions that $\phi$ goes asymptotically to the false vacuum $\phi_F$
at infinity and is ``close'' to the true vacuum at the origin. To ensure the regularity of the solution,
the additional condition $d \phi/dr = 0$ at the origin is also needed, where $r$ is the Euclidean distance.

Solving the previous equation requires non-trivial numerical algorithms. The problem can be addressed analytically
in the thin-wall limit, which corresponds to a first-order phase transition with weak supercooling.
In this case the height of the barrier between the two minima, the true $\phi_T$ and the false $\phi_F$ ones,
is much larger than their relative depth $\epsilon = V(\phi_T, T) - V(\phi_F, T)$. The thin-wall approximation, however,
is not always justified in our parameter space. For this reason the analytic approach can only be used to derive
approximate results that capture the overall dependence of the properties of the phase transition
on the parameters of the model. We report in appendix~\ref{app:thinwall} the results obtained through this approach.

In this section, instead, we present the exact results obtained through a numerical integration of the bounce equations.
We performed the computation with the dedicated package \texttt{CosmoTransitions}~\cite{Wainwright:2011kj}.
We also checked independently through our own code that the numerical results we found are stable and
reliable.\footnote{Our code has been implemented in Mathematica~\cite{Mathematica} and is based on a variation of the
algorithm described in ref.~\cite{Athron:2019nbd}, which relies on the linearisation of the bounce equations.}

\bigskip

The first issue we need to address is to check in which regions of the parameter space the EWSB transition actually takes place.
After the appearance of two degenerate minima at $T=T_c$, as the Universe cools down the ratio of bubble formation, $\Gamma \sim e^{-S_3/T}$, increases. Successful nucleation starts when this rate eventually balances the Hubble expansion. However, if the bounce action is too large, it can happen that the balancing condition is never reached and
the system remains trapped in the metastable vacuum.
This can be the case if the portal coupling $\lambda_{h\eta}$ is too large, producing a high barrier between the two minima.

\begin{figure}
\centering
{\includegraphics[width=0.43\textwidth]{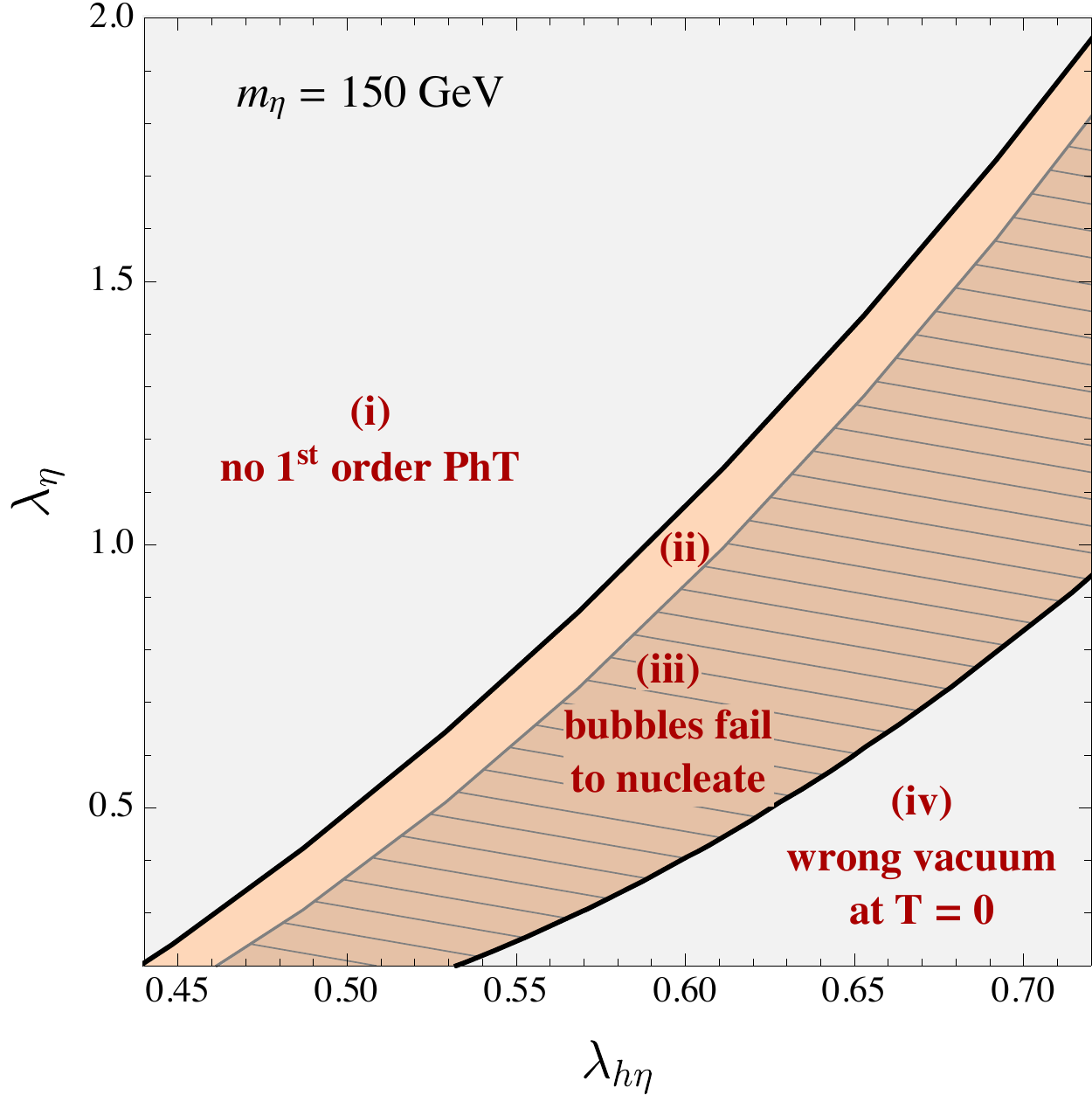}}\hspace{2em}
{\includegraphics[width=0.43\textwidth]{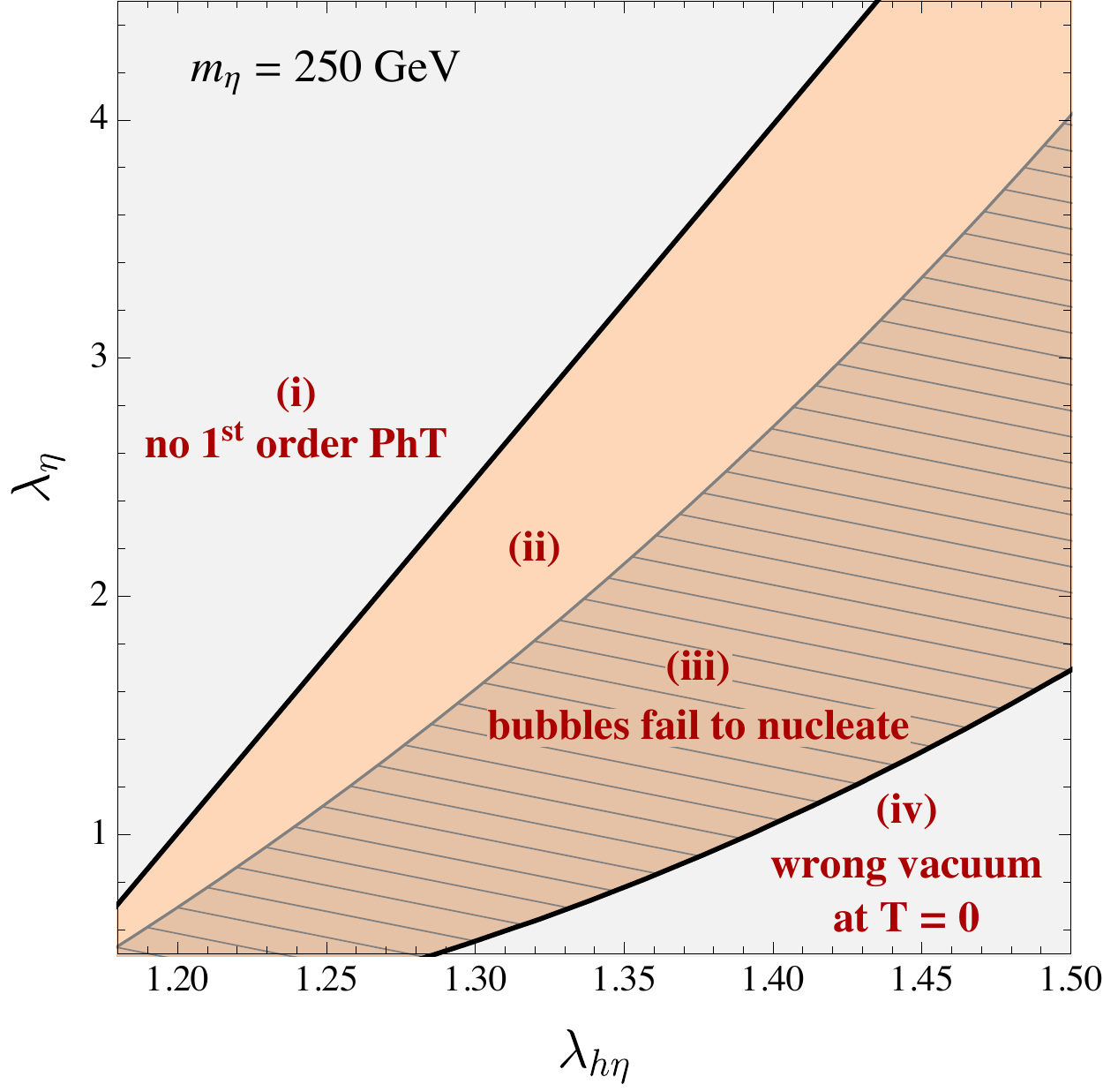}}
\caption{Phase transition diagram, for fixed singlet scalar mass $m_\eta = 150\,{\rm GeV}$ (left panel) and $m_\eta = 250\,{\rm GeV}$ (right panel). The thermal history of the models in the various regions is explained in the text.\label{Fig:EWPhTdiagram}}
\end{figure} 

We show in fig.~\ref{Fig:EWPhTdiagram} the phase transition diagram for two choices of the mass of the singlet scalar,
namely $150\,{\rm GeV}$ and $250\,{\rm GeV}$. The results are shown as a function of the two free parameters,
$\lambda_\eta$ and $\lambda_{h\eta}$.
For a given value of the quartic $\lambda_\eta$, the portal coupling $\lambda_{h\eta}$ determines the thermal history of the model. A few different scenarios can happen:
\begin{itemize}
\item[(i)] For small values of $\lambda_{h\eta}$, in the grey region in the upper left corner, the EWSB minimum is always the global minimum of the potential.
This can happen either when no additional minimum is present (towards the upper left part of the diagram), or when a
local minimum $(0, w)$ is present but does not become deeper than the EWSB one at $T \neq 0$.
In this region a second-order transition from the symmetric phase to the EWSB one takes place.
\item[(ii)] For larger values of $\lambda_{h\eta}$, the additional vacuum $(0, w)$ is a local minimum
at $T=0$, but becomes degenerate with the EWSB one at a critical temperature $T_c$. If the
bounce action is small enough, as it happens in the orange region, nucleation can occur, and a first-order phase transition
connects the $(0, w)$ vacuum to the EWSB one. This region gives rise to the two-step phase transition we are interested in.
\item[(iii)] If the bounce action is too large, as it happens in the dark orange striped region, although the two minima become degenerate,
nucleation is never efficient. In this case the system remains trapped in the metastable vacuum $(0, w)$ and no EWSB occurs.
\item[(iv)] Finally for large values of $\lambda_{h\eta}$, in the grey region in the lower right corner, the global minimum at $T=0$ is the
$(0, w)$ point and not the EW vacuum. These configurations never give rise to EWSB and are clearly not viable.
\end{itemize}
From fig.~\ref{Fig:EWPhTdiagram}, one can see that the region of parameter space with a viable two-step transition
tends to become wider for larger singlet masses. Although configurations with two vacua are easy to realise in these
models (full orange regions (ii) and (iii) in the plots), the condition of successful bubble nucleation becomes harder to satisfy
for smaller $m_\eta$. For instance, we found that for $m_\eta = 100\,{\rm GeV}$ only a very narrow strip
of parameter space allows for nucleation. In general, for $m_\eta < m_h$, a first order phase transition can be realised only for suitably chosen values of $\lambda_\eta$.
We also checked the results presented in ref.~\cite{Kurup:2017dzf} and found quantitative agreement with the $m_\eta = 300\,{\rm GeV}$ benchmark point considered in that paper.

\bigskip

\begin{figure}[t]
\centering
{\includegraphics[width=0.43\textwidth]{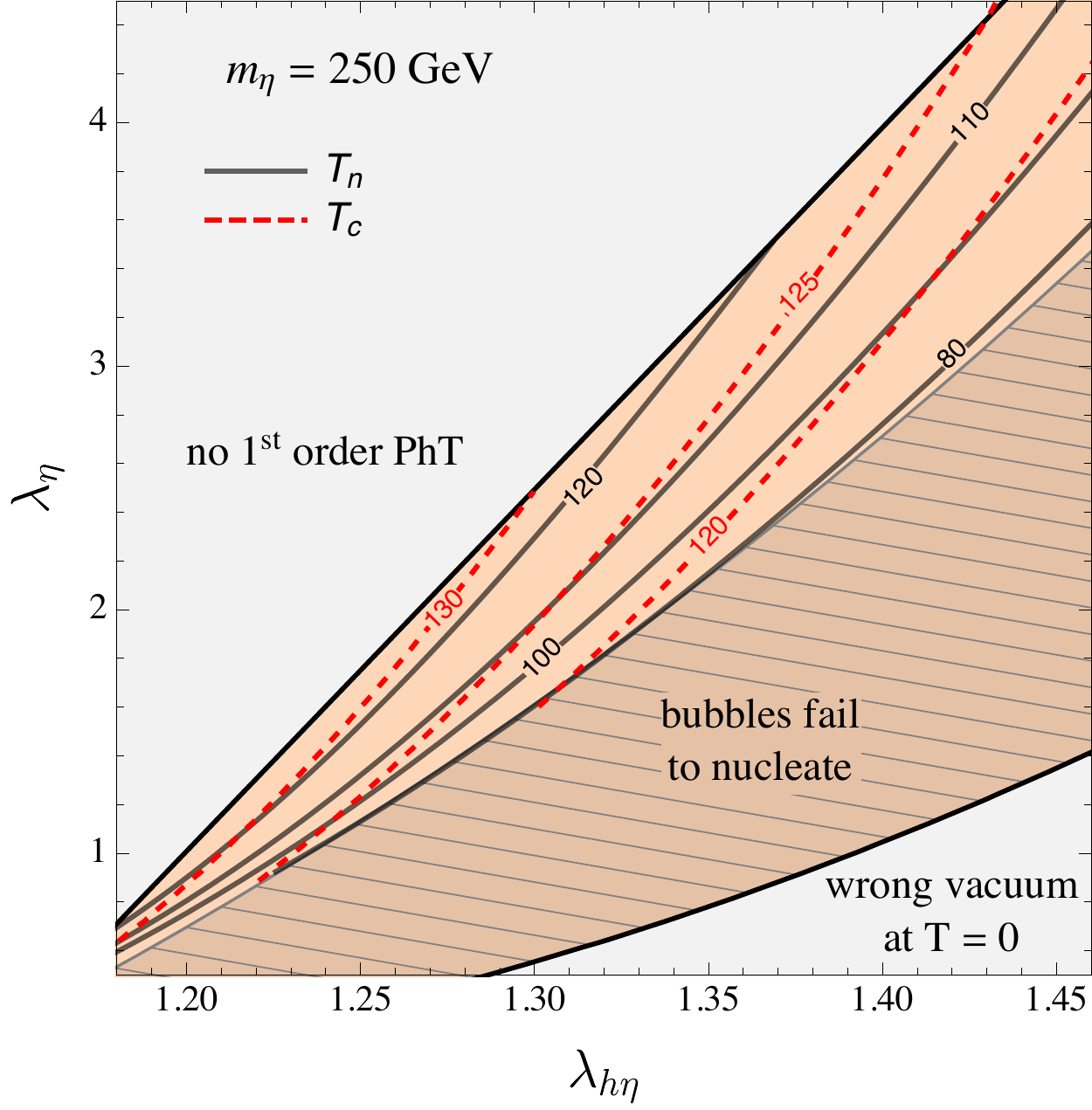}}\hspace{2em}
{\includegraphics[width=0.43\textwidth]{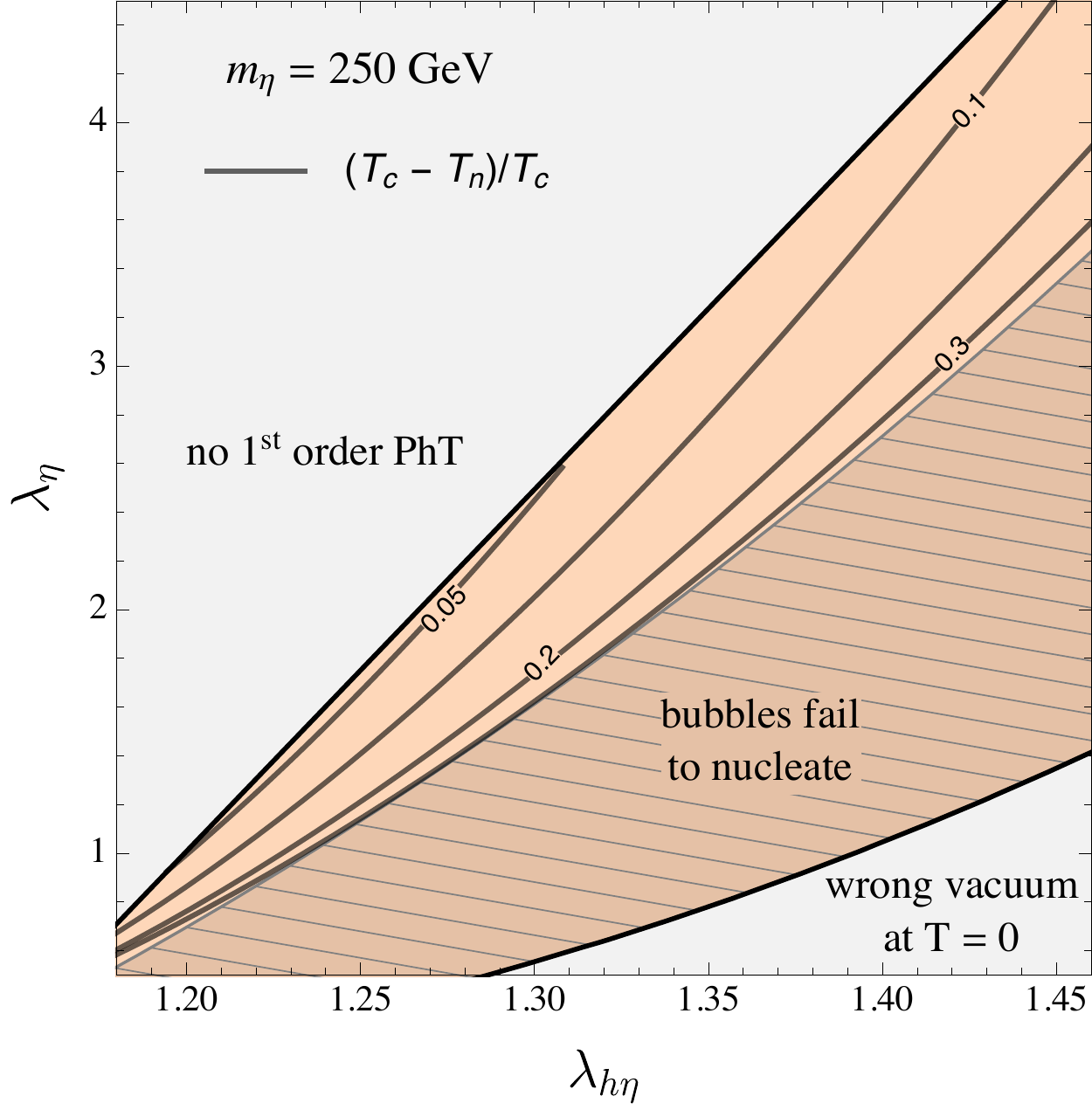}}
\caption{Left panel: Critical temperature $T_c$ and nucleation temperature $T_n$. Right panel: amount of supercooling. \label{fig:PhT_Temperature}}
\end{figure}

We can now study the properties of the first-order phase transition. For definiteness we focus on the benchmark with
$m_\eta = 250\,{\rm GeV}$.
In fig.~\ref{fig:PhT_Temperature} (left) we show the two temperatures that characterise the phase transition, namely
the critical temperature $T_c$ and the nucleation temperature $T_n$. One can see that both $T_c$ and $T_n$ decrease
for larger values of $\lambda_{h\eta}$. The critical temperature varies in the range $115\,{\rm GeV} \lesssim T_c \lesssim 130\,{\rm GeV}$, while the nucleation temperature is in the range $80\,{\rm GeV} \lesssim T_c \lesssim 125\,{\rm GeV}$.
In the right panel of the fig.~\ref{fig:PhT_Temperature} we show the amount of supercooling, namely $(T_c - T_n)/T_c$. One can see that for small values of $\lambda_{h\eta}$ there is almost no supercooling, whereas for a larger values it is possible
to achieve an amount of supercooling of order $30\%$.
As we will see in the following section, the region of the parameter space with smaller $T_n$ and higher supercooling
can be more easily probed by future gravitational wave experiments, since the peak of the corresponding
wave spectrum moves, as $T_n$ decreases, towards frequencies where they have the maximum sensitivity.  

\begin{figure}[t]
\centering
{\includegraphics[width=0.43\textwidth]{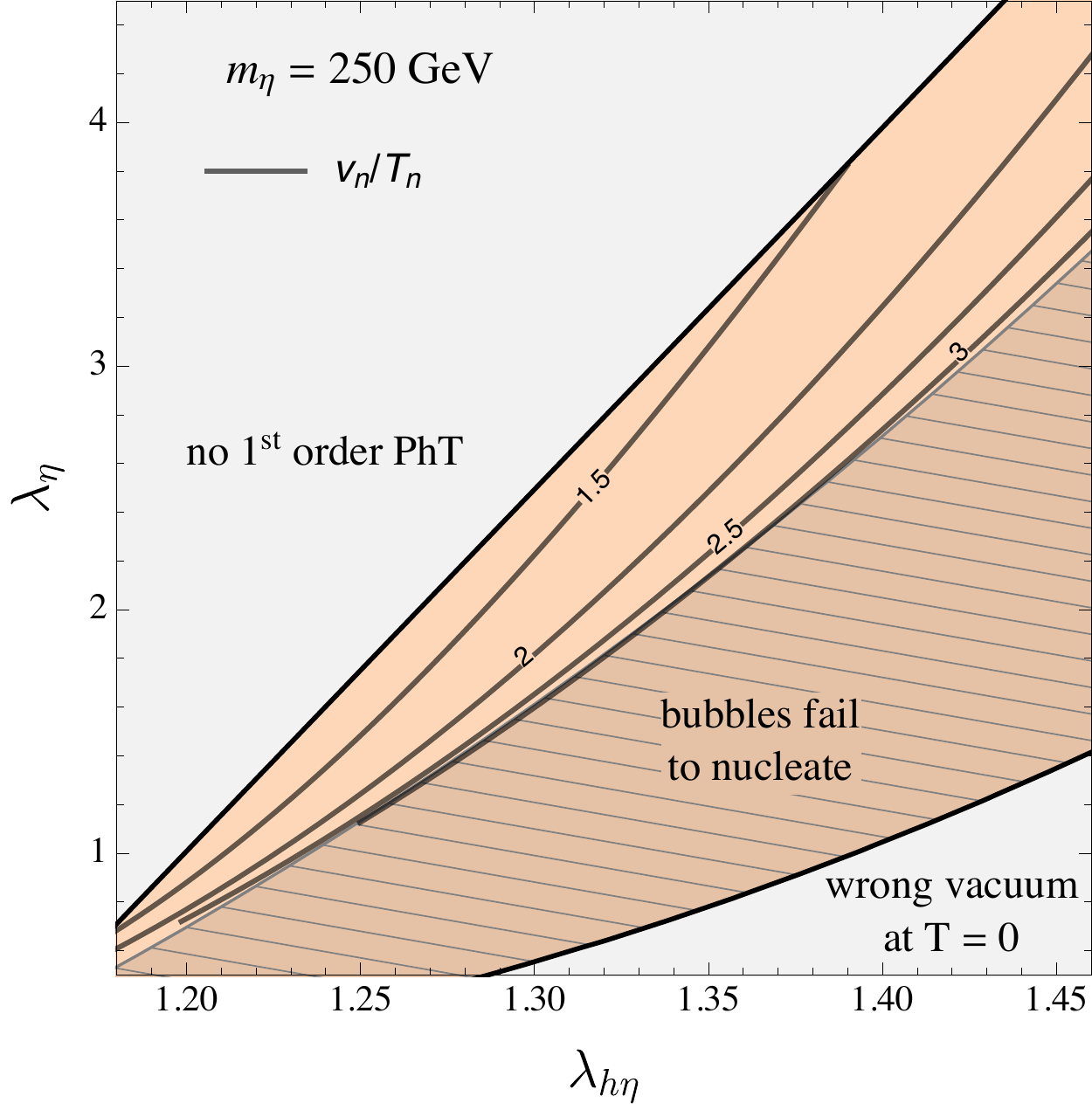}}\hfill
\raisebox{2.em}{\includegraphics[width=0.55\textwidth]{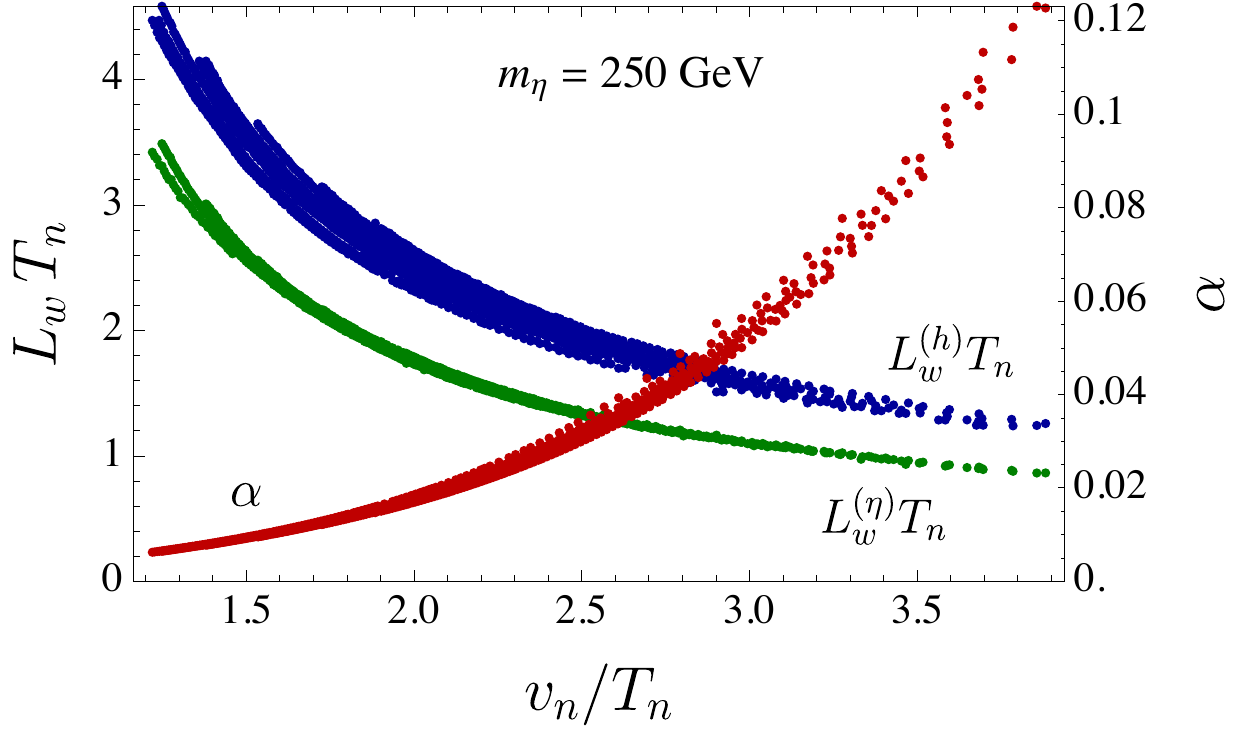}}
\caption{Left panel: Strength of the phase transition $v_n/T_n$. Right panel: Scatter plot of the vacuum energy density parameter
$\alpha$ (red dots) and of the bubble width $L_w T_n$ for the Higgs (blue dots) and the $\eta$ (green dots) components
as a function of the phase transition strength $v_n/T_n$. \label{fig:PhT_Strength}}
\end{figure}

In fig.~\ref{fig:PhT_Strength} (left) we show the strength $v_n/T_n$ of the phase transition, a crucial parameter for the EWBG, which increases, as expected, with the portal coupling. 
For $v_n/T_n \gtrsim 1$, the EWPhT is strongly first order and prevents the EW sphaleron processes inside the broken phase to washout the baryon asymmetry generated in front of the bubble wall.

Another important parameter is the ratio $\alpha = \rho_\textrm{vac}/ \rho_\textrm{rad}$ of the vacuum energy density $\rho_\textrm{vac}$, released to the primordial plasma during the transition, and the critical energy $\rho_\textrm{rad}$ at the transition temperature. This parameter controls the size of the signal of gravitational waves from EWPhT and represents a measure of the strength of the phase transition.
In fig.~\ref{fig:PhT_Strength} (right) we show a scatter plot of the values of $\alpha$ (red dots) as a function of the phase transition strength. One can see that a strong correlation between the two quantities is present. For small $v_n/T_n$ one
can show that $\alpha \propto (v_n/T_n)^2$ (see eq.~(\ref{eq:alpha_vc}) in appendix~\ref{app:thinwall}),
however this relation receives order-one corrections for $v_n/T_n \gtrsim 2.5$.
On the same figure we also show the scatter plot for the width of the bubble wall $L_w$, which is reported in the combination $L_w T_n$ both for the Higgs (blue dots) and the $\eta$ (green dots) components.
Also in this case a strong correlation with the strength of the phase transition is present.

\begin{figure}[t]
\centering
{\includegraphics[width=0.43\textwidth]{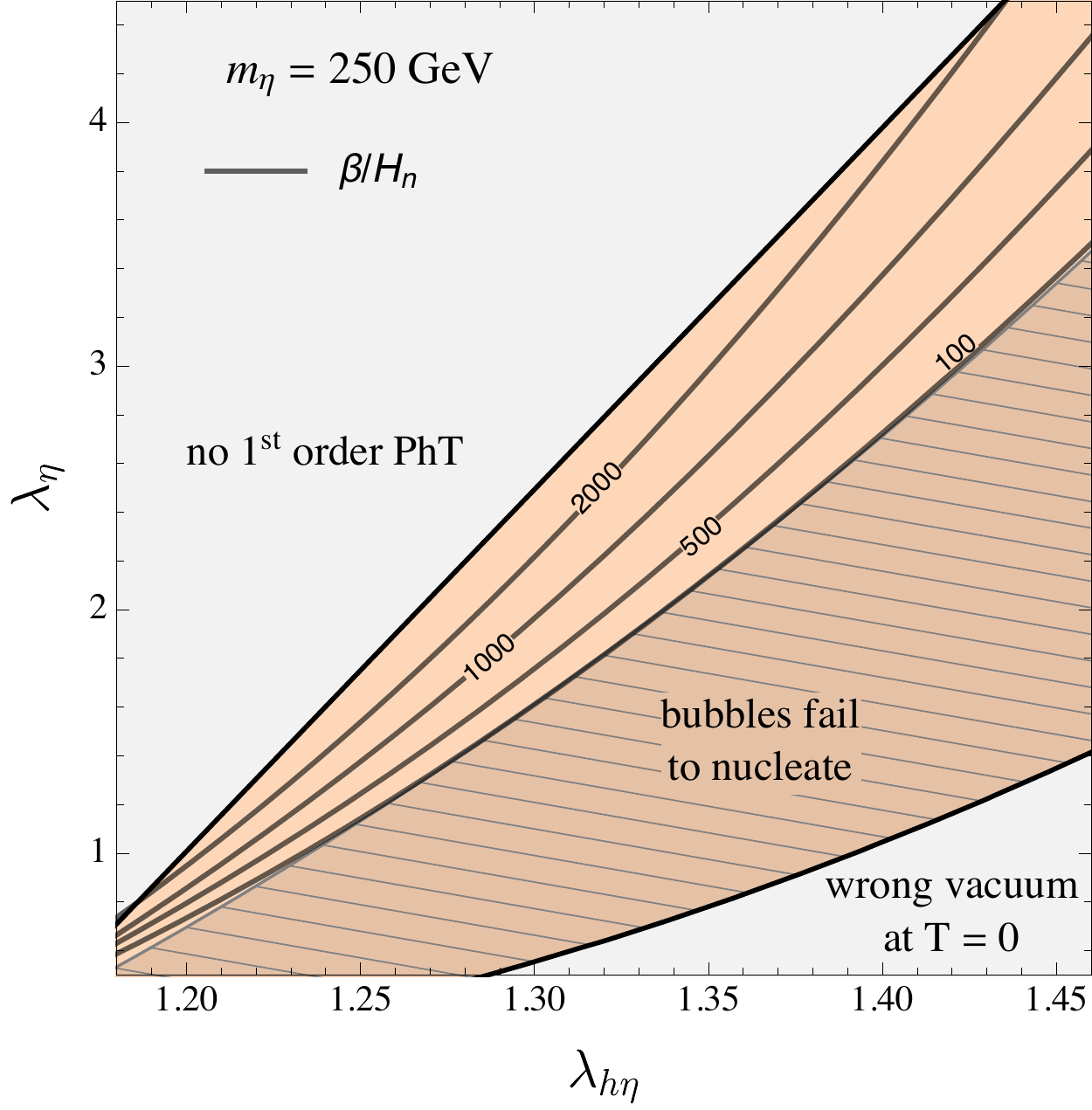}}\hfill
\raisebox{2.em}{\includegraphics[width=0.53\textwidth]{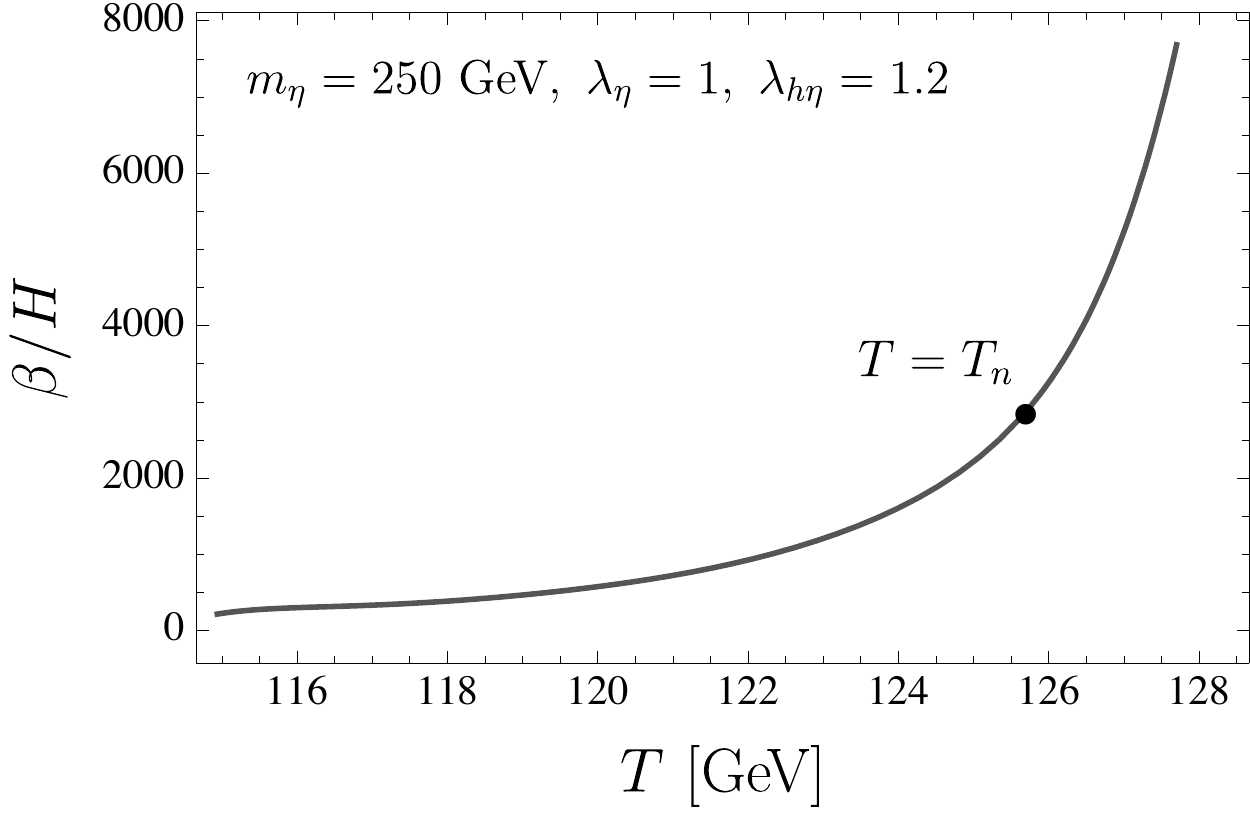}}
\caption{Left panel: Inverse time duration $\beta/H_n$ of the phase transition. Right panel: Dependence of
$\beta/H_n$ on the transition temperature.\label{fig:PhT_beta}}
\end{figure}

The last parameter we consider is the inverse time duration of the phase transition, normalised to the Hubble rate.
This quantity controls the amplitude of the gravitational wave spectrum and can be computed from the variation of the bounce action with respect to the temperature
\bea
\frac{\beta}{H_n} = T \frac{d}{d T}\left( \frac{S_3}{T} \right)\bigg |_{T_n} \,.
\eea
The numerical results for $\beta/H_n$ are shown in the left panel of fig.~\ref{fig:PhT_beta}.
Larger values for $\beta/H_n$ ($\beta/H_n \sim 3000$) are obtained for small $\lambda_{h\eta}$, i.e.~for larger
phase transition temperatures. On the other hand, for larger $\lambda_{h\eta}$, the values of $\beta/H_n$
are significantly smaller ($\beta/H_n \sim 100$).
It must be noticed that the value of $\beta/H_n$ strongly depends on the transition temperature. As can be seen
in the right panel of fig.~\ref{fig:PhT_beta} for a benchmark point,
even a $\textit{few}\;{\rm GeV}$ difference in the phase transition temperature can modify $\beta/H_n$ by almost
one order of magnitude.

\section{Gravitational waves}
\label{sec:gw}

The transition between two minima separated by a potential barrier is described by the nucleation of bubbles of true vacuum in the background of metastable vacuum.
The bubbles expand, collide and eventually coalesce filling the whole space. This phenomenon is characterised by a huge release of energy which propagates, in part, through GWs.
Since bubbles collide incoherently in different regions of space, the corresponding signal is a stochastic background of GWs.
The peak of the frequency spectrum for phase transitions at the electroweak scale typically lies in the sensitivity range of future experiments, such as the European space-based interferometer LISA~\cite{Seoane:2013qna,Klein:2015hvg,Audley:2017drz}, BBO~\cite{Harry:2006fi,Yagi:2011wg,Crowder:2005nr} and the proposed Japanese detector DECIGO~\cite{Kawamura:2006up}.

Three different mechanisms of GW production are at work during bubble nucleation: bubble collision~\cite{Kosowsky:1991ua,Kosowsky:1992rz,Kosowsky:1992vn,Kamionkowski:1993fg,Huber:2008hg,Caprini:2007xq}, sound waves in the plasma after the collision~\cite{Hindmarsh:2013xza,Giblin:2013kea,Giblin:2014qia,Hindmarsh:2015qta} and magnetohydrodynamic turbulence effects in the plasma~\cite{Caprini:2006jb,Kahniashvili:2008pf,Kahniashvili:2008pe,Kahniashvili:2009mf,Caprini:2009yp}. The three contributions can be approximately combined linearly 
\bea
h^2 \Omega_\textrm{GW} \simeq h^2 \Omega_\phi + h^2 \Omega_\textrm{SW} + h^2 \Omega_\textrm{MHD}.
\eea
The amplitude of the spectrum and the position of the frequency peak are mainly characterised by the nucleation temperature ($T_n$), the vacuum energy, normalised to the critical energy, released to the primordial plasma during the transition ($\alpha$), and the duration of the transition itself ($\beta$). 
These parameters are supplemented by the efficiency coefficients ($\kappa$) and the bubble velocity ($v_w$).
The analytic formulas for the different contributions to the GW spectrum are given in ref.~\cite{Caprini:2015zlo}.

The bubble velocity is the result of the balance between the force driving the expansion of the bubble and its friction with the plasma. 
In particular, three different regimes can be identified depending on $v_w$ compared to the sound velocity in the plasma $v_s$ (hybrid possibilities can also be present): deflagration ($v_w < v_s < 1$), detonation ($v_s < v_w < 1$) and runaway regime ($v_w = 1$). In the first two cases, the bubble velocity reaches a constant value because the interactions of the bubble surface with the particles in the plasma can balance the expansion. On the contrary, in the runaway case the pressure driving the bubble expansion overcomes the friction and leads to an indefinite velocity growth. 
The bubble velocity represents a crucial parameter since an efficient production of baryon asymmetry prefers the deflagration regime while the observability of GWs is more favourable in the detonation and runaway scenarios. 
It has been shown recently \cite{Dorsch:2016nrg}, in the context of a two step phase transition driven by the extra scalar state of a second Higgs doublet, that in the region of parameter space where the EW baryogenesis is achievable, the GW spectrum of the EWPhT is within the sensitivity reach of future interferometers. Indeed, even for very strong phase transitions, $v_n/T_n \simeq 4$, the bubble wall velocity remains subsonic. The determination of $v_w$ is very challenging and requires the microscopic calculation of the friction term and the solution of the Boltzmann equations modelling the interaction of the scalar fields with the thermal plasma, see for instance refs.~\cite{Bodeker:2009qy,Bodeker:2017cim,Espinosa:2010hh,Konstandin:2014zta,Kozaczuk:2015owa,Dorsch:2018pat}. The exact computation of the velocity is beyond the scope of this work, here we use for the sake of simplicity the prediction of $v_w$, as a function of $\alpha$, that has been estimated in ref.~\cite{Dorsch:2016nrg}.

\begin{figure}
\centering
{\includegraphics[width=0.48\textwidth]{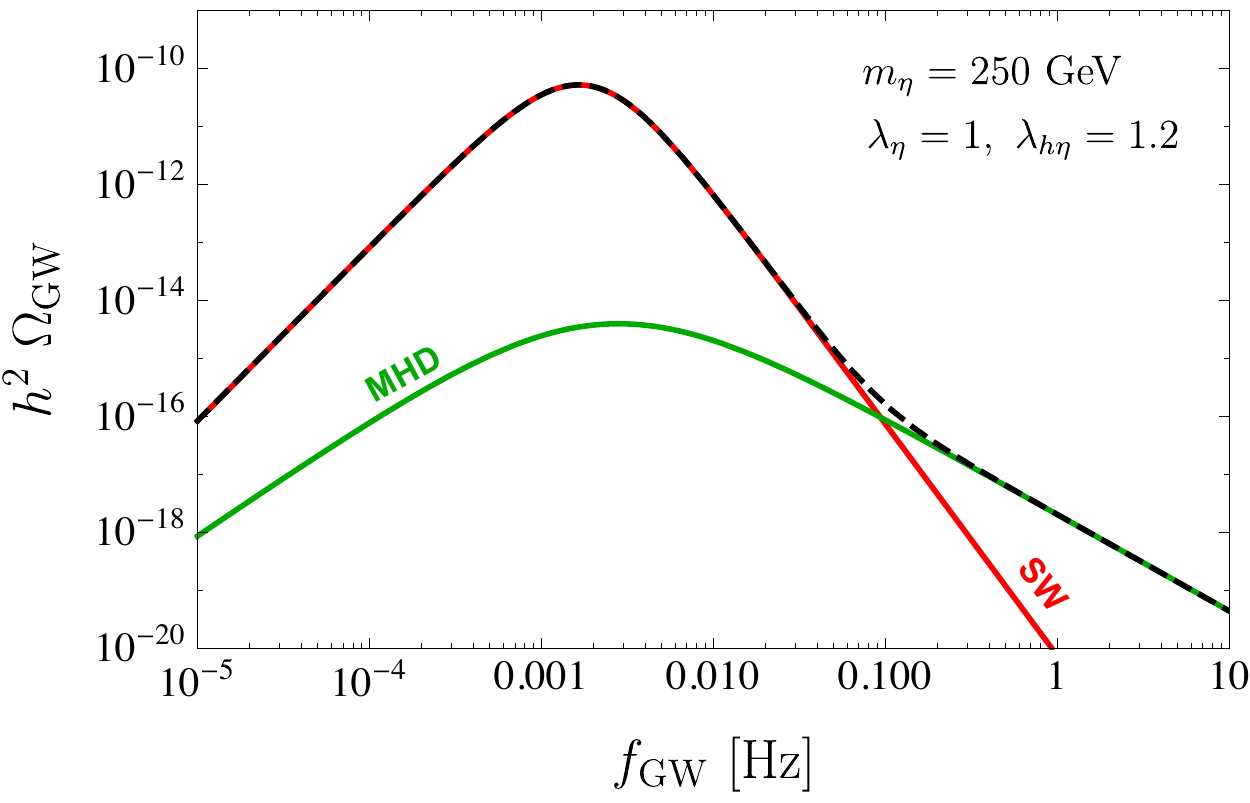}}\hfill
{\includegraphics[width=0.48\textwidth]{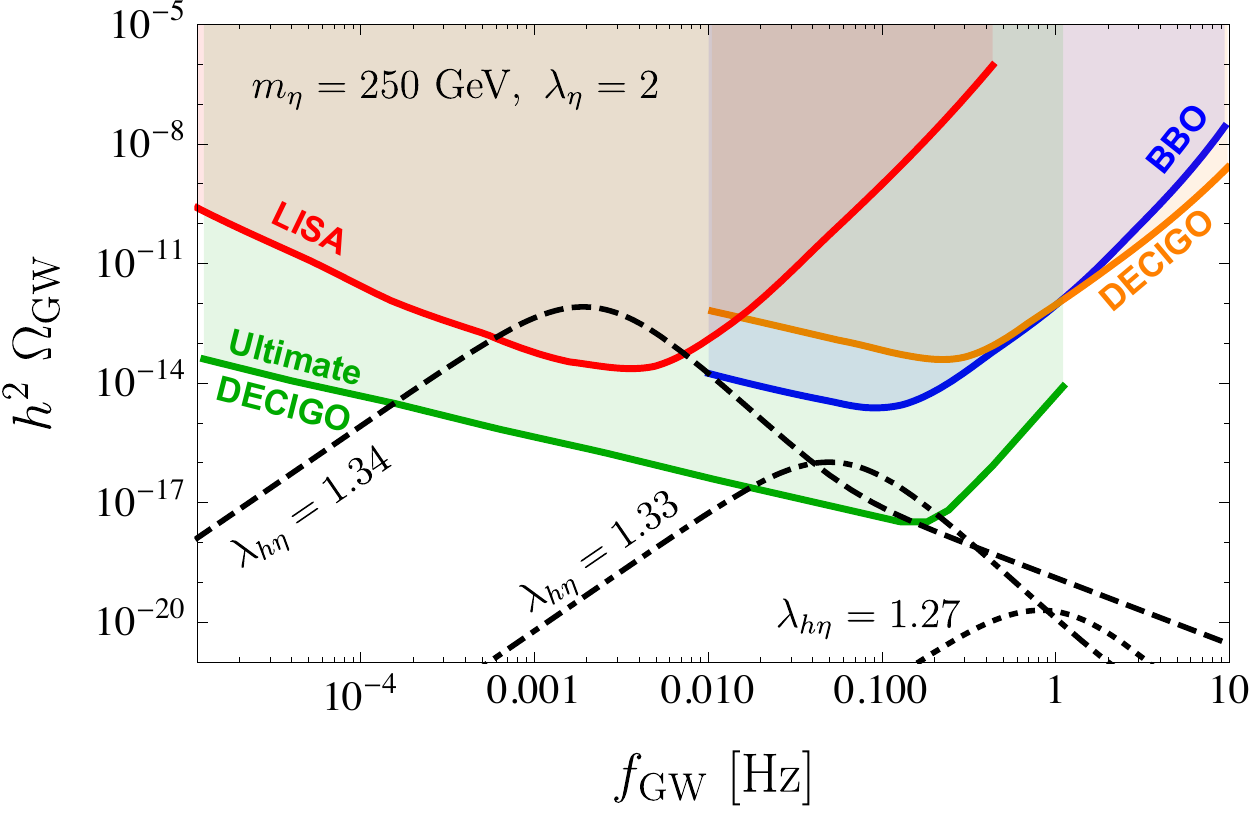}}
\caption{ Left panel: Leading contributions to the GW spectrum in the non-runaway regime for the benchmark point $m_\eta = 250$ GeV, $\lambda_{h\eta} = 1.4$ and $\lambda_\eta = 3$. 
Red, green and dashed lines correspond, respectively, to GWs from sound waves in the plasma, magnetohydrodynamic turbulences and the linear combinations of the two.
Right panel: GW spectra as a function of the frequency for three benchmark points with $m_\eta = 250$ GeV, $\lambda_\eta = 2$ and $\lambda_{h\eta} = 1.27$ (dotted), $\lambda_{h\eta} = 1.33$ (dot-dashed), $\lambda_{h\eta} = 1.34$ (dashed). Sensitivity curves of some future space-base interferometers are also shown.
\label{GWspetrcum}}
\end{figure} 

The three sources of GW are characterised by different peak frequencies that, if sufficiently separated, can lead to a non-trivial structure for the spectrum, helping in the extraction of the signal from the instrumental background noise. As an example, we show in fig.~\ref{GWspetrcum} (left) the contribution of the different components to $h^2 \Omega_\textrm{GW}$ for a selected point with $m_\eta = 250$ GeV, $\lambda_{h\eta} = 1.4$ and $\lambda_\eta = 3$. Notice that in the non-runaway regime the contributions from bubble collisions can be neglected. Numerical simulations show that the relative distance between the peaks of the two spectra is fixed, $f^\textrm{peak}_\textrm{SW}/f^\textrm{peak}_\textrm{MHD} \simeq 0.7$, and that the signal from sound waves decays faster for larger GW frequency $f_\textrm{GW}$, namely $h^2 \Omega_\textrm{SW} \sim f_\textrm{GW}^{-4}$ and $h^2 \Omega_\textrm{MHD} \sim f_\textrm{GW}^{-5/3}$. This explains the typical shoulder of the GW spectrum at high frequencies.

In fig.~\ref{GWspetrcum} (right) we show the sensitivity reach of the three future GW experiments LISA, BBO and DECIGO, as well as the prediction of the GW spectra for three benchmark points. The benchmarks have fixed $m_\eta = 250$ GeV and $\lambda_\eta = 2$ and are defined, respectively, by $\lambda_{h\eta} = 1.27$ (dotted line), $\lambda_{h\eta} = 1.33$ (dot-dashed line) and $\lambda_{h\eta} = 1.34$ (dashed line). The values of $\lambda_{h\eta}$ have been chosen to span the region of successful phase transition pattern for $\lambda_\eta = 2$ (see right panel of fig.~\ref{Fig:EWPhTdiagram}). As $\lambda_{h\eta}$ increases, the GW signal strengthens and the peak of the spectrum shifts towards smaller frequencies, which are preferred by space-based interferometers. Indeed, the frequency peak
for the sound-wave (SW) and magnetohydrodynamic turbulence (MHD) components is given by
\bea
f^\textrm{peak}_\textrm{SW  (MHD)} = 1.9 \,(2.7) \times 10^{-5} \, \textrm{Hz} \frac{1}{v_w} \, \left( \frac{\beta}{H_n} \right) \left( \frac{T_n}{ 100 \, \textrm{GeV}} \right) \left( \frac{g_*}{100} \right)^{\frac{1}{6}} \,,
\eea
where $g_*$ is the number of relativistic degrees of freedom in the plasma at the time of the phase transition. It scales linearly with $\beta/H_n$ and $T_n$, which both decrease when the portal coupling increases.

\begin{figure}
\centering
{\includegraphics[width=0.45\textwidth]{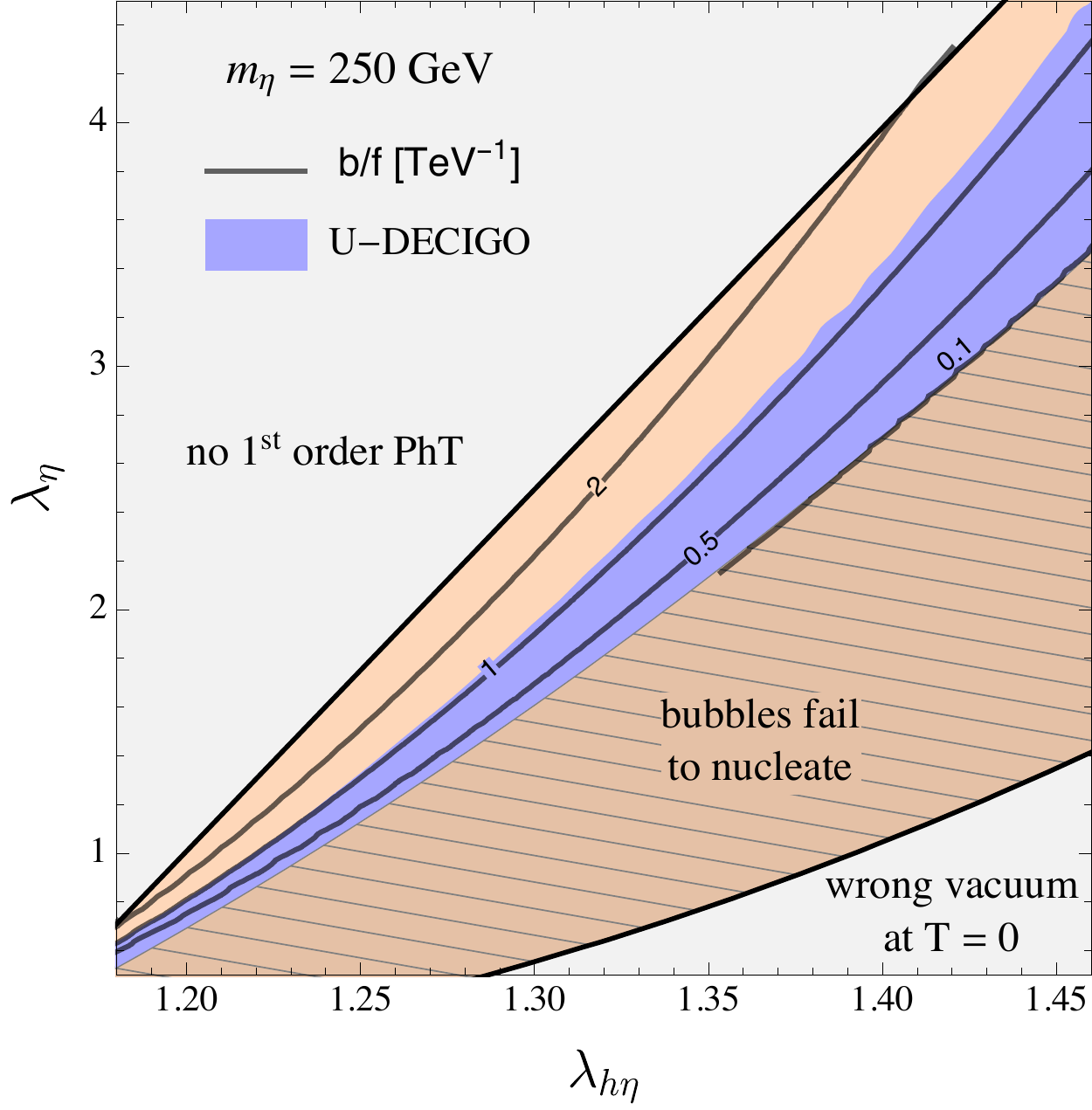}}
\caption{Observational reach of the gravitational signal from the first order EWPhT at Ultimate-DECIGO for a benchmark
scenario with $m_\eta = 250\,{\rm GeV}$. The solid grey contours show the values of $b/f$ needed to guarantee
a sufficient amount of CP violation to achieve EW baryogenesis.
\label{fig:GW_reach}}
\end{figure}

The prospect of observations of GWs at Ultimate-DECIGO in the two dimensional parameter space of $\lambda_{h\eta}$ and $\lambda_{\eta}$ for singlet mass $m_\eta = 250\,{\rm GeV}$ is depicted in fig.~\ref{fig:GW_reach}. We decided not to show the region accessible at LISA, since it can only test a narrow strip at the right edge of the two-step transition region.

\section{Electroweak baryogenesis}\label{sec:EW_baryo}

The out-of-equilibrium dynamics provided by the first order EWPhT fulfils only one of the three Sakharov's conditions required to realise baryogenesis. 
A sufficiently strong source of CP violation is also needed in order to trigger an asymmetry between matter and antimatter. 

In principle, additional sources of CP violation have to be expected in CHMs due to the presence of additional complex phases
(for instance in the elementary--composite mixing parameters). Some restrictions on the amount of CP violation might
be present if we want to ensure the $P_\eta$ invariance of the scalar potential. In fact, as we discussed, this requirement
typically obliges the composite sector to be invariant under CP.

However, an additional source of CP violation is typically present as a consequence
of the non-linear dynamics of the Goldstones. This relies on the presence of the dimension-$5$ operator
$\eta \, h \, \bar t_L  t_R$, which can have a complex coefficient and is naturally present in most of the models based
on the $\SO(6)/\SO(5)$ coset. Indeed we saw that such operator is present in the $({\bf 6}, {\bf 6})$
and $({\bf 6}, {\bf 20}')$ scenarios.

At $T = 0$, in the EWSB vacuum, the $s \, h \, \bar t_L  t_R$ operator gives rise to small
CP violating effects, which can be compatible with the present constraints (we will discuss this aspect at the end of this section).
Moreover a possible complex phase in the top mass can always be rotated away through a redefinition of the top
field and is thus unphysical.
On the contrary, when the both the Higgs and the singlet get a VEV, a new complex phase is induced in the top mass.
This obviously happens in the bubble walls during the EWSB phase transition. 
Since the Higgs and the singlet VEVs are space dependent, the new phase in the top mass cannot be reabsorbed by a redefinition of the fermionic fields and provides a new source of CP violation that can trigger EWBG.

The phase in the top mass $\Theta_t$ can be defined as
\begin{equation}
m_t(r) = |m_t(r)| e^{i \Theta_t(r)}
\end{equation}
with $r$ denoting the direction perpendicular to the bubble wall. 
For each of the scenarios discussed previously, the complex phase can be extracted from the $\mathcal O_t$ operators that give rise to the top Yukawa. To be as general as possible, we rewrite them here as
\begin{equation}\label{eq:top_eta_coupling}
\mathcal O_t = y_t \left(1 + i \frac{b}{f} \eta  \right) \frac{h}{\sqrt{2}} \, \bar t_L t_R + \textrm{h.c.}\,.
\end{equation}
The phase of top quark mass is then given by
\begin{equation}
\Theta_t(r) = \arctan \left(b \frac{w(r)}{f} \right) 
\end{equation}
with $w(r)$ exhibiting the usual kink profile along the $r$ direction. 
The coefficient $b$ is determined by the particular fermion embedding. 
For instance, in the $(\bold 6,\bold 6)$ case $b = \tan \theta_{u_6}$ is completely fixed by the admixture of $t_R$ embedding in the 5-th and 6-th components of the fundamental of $\SO(6)$. In the $(\bold{15}, \bold 6)$ case, instead, $b$ vanishes identically. Therefore, while this model can provide a first order EWPhT, it is not suitable for the EWBG unless one allows for explicit CP-violating interactions. Finally, in the $(\bold 6, \bold{20}')$ scenario, two independent operators contribute to the top mass, and $b$ is determined by the ratio of their coefficients as well as by the two angles parametrising the $t_R$ embedding in the $\bold{20}'$ multiplet.

In the semi-classical approximation, the complex phase in the top quark mass induces different dispersion relations for particles and antiparticles, which, through the electroweak sphaleron processes in the symmetric phase, can trigger a net baryon asymmetry \cite{Fromme:2006wx}. The latter is preserved only if the same sphaleron processes are quickly dumped in the broken domain, namely if the phase transition is sufficiently strong. 

The baryon asymmetry depends linearly on the variation of the phase of the top mass and non-trivially on the dimensionless combinations $v_n/T_n$ and $L_w T_n$. In particular, it increases with increasing strength and decreasing $L_w$. On the other hand, the dependence on the bubble wall velocity $v_w$ is expected to be mild as long as the deflagration regime is concerned \cite{Fromme:2006cm}.  
Rather than numerically solving the system of transport equations \cite{Bodeker:2004ws}, we use the results obtained in ref.~\cite{Espinosa:2011eu} which we recast in fig.~\ref{fig:GW_reach} in the plane of the two quartic couplings for our benchmark singlet mass $m_\eta = 250\,{\rm GeV}$. In fig.~\ref{fig:GW_reach} we show the size of $b/f$ required to successfully reproduce the observed baryon asymmetry, $(n_B - n_{\bar B})/n_\gamma \simeq 6 \times 10^{-10}$.
One can see that values of $b/f$ of ${\cal O}(1/{\rm TeV})$ are sufficient to generate a realistic asymmetry. In some regions
of the parameter space we even need significantly smaller values, $b/f \sim 0.1\,{\rm TeV}$, which can be easily
realised in the models we considered.

\bigskip

As well known, EWBG is more efficient for subsonic bubble walls, since the CP violating interactions have more time to generate a particle/antiparticle asymmetry in front of the wall which is then converted into baryon asymmetry by the electroweak sphalerons. In this regime, the scalar field component of the GW spectrum from bubble collision is strongly suppressed. 
Nevertheless, as shown in the previous section, the contributions from sound waves and turbulence effects in the plasma leave open the intriguing possibility to detect GW signals in the same region of parameter space in which EWBG is possible. Indeed, as confirmed in ref.~\cite{Dorsch:2016nrg}, subsonic wall velocity can be compatible with sufficiently strong phase transitions. 
For $b/f \lesssim \textrm{TeV}^{-1}$, the CHM can explain the observed matter-antimatter asymmetry and provides, at the same time, GW signals potentially detectable with the future generations of interferometers.  

\bigskip

To conclude we consider the constraints on the amount of CP violation coming from the experimental data.
The scenarios we considered are characterised by spontaneous CP-violation driven by the breaking of the $P_\eta$ parity through thermal effects.
However, as shown in refs.~\cite{McDonald:1995hp,Espinosa:2011eu}, a small explicit breaking of $P_\eta$ is also needed
to bias the population of one of the two $(0, \pm w_c)$ configurations which arise in the two-step process after
the first second-order phase transition.\footnote{A possible source of explicit $P_\eta$ breaking can be provided by the embedding of the light-generation fermions.}
If this were not case, a net baryon asymmetry would be evened out
by a net antibaryon asymmetry in different patches of the Universe. 
By parametrising the explicit $P_\eta$ breaking with $\Delta V = V(w_c) - V(-w_c)$, one can find that its lower bound is very weak $\Delta V/T^4 \gg H/T \sim 10^{-16}$~\cite{McDonald:1995hp,Espinosa:2011eu} and that $\Delta V$ can be taken small enough not to affect the previous analysis.

On the other hand, an explicit $P_\eta$ breaking term would also induce a mixing between the Higgs
and the CP-odd singlet $\eta$. This mixing, together with the CP-violating $\eta \, h \, \bar t_L  t_R$ operator, can give rise
to contributions to the electron and light quark electric dipole moment (EDM).
These effects originate at the two-loop level through Barr--Zee-type diagrams~\cite{Barr:1990vd} involving a top loop.
Their size is given by
\begin{equation}
\frac{d_e}{e} \simeq \frac{\sin 2\phi}{2} \frac{32}{3} \frac{\alpha_\textsc{em}}{(4 \pi)^3} \frac{m_e}{v}   \frac{b}{f}  \left[ g\left(\frac{m_t^2}{m_h^2} \right) - g\left(\frac{m_t^2}{m_\eta^2} \right) \right]\,,
\end{equation}
where $\phi$ is the mixing angle between the Higgs and the singlet and the function $g(x)$ can be found in ref.~\cite{Barr:1990vd}. The strongest constraint arises from the bound on the electron EDM~\cite{Andreev:2018ayy}.
Assuming a small mixing angle and $m_\eta =  250$ GeV, this translates into
\begin{equation}
\left|\phi \frac{b}{f} \right| \lesssim 1.3 \times 10^{-2} \,\, \textrm{TeV}^{-1}\,,
\end{equation}
which can be easily satisfied by a sufficiently small $P_\eta$ breaking.\footnote{Notice that in CHMs
additional CP-violating interactions of the Higgs with the top and the top partners are typically present,
which translate into strong bounds on the mass of the top partners and on their couplings~\cite{Panico:2017vlk,Panico:2018hal}. These effects, however
are independent with respect to the dynamics we are interested in for EWBG, so we can neglect them in our analysis.}

If the singlet $\eta$ is also coupled to the electron or to the light quarks through operators of the form
$\eta \, h \, \bar \psi_L \psi_R$, additional contributions to the EDMs arise through Barr--Zee diagrams involving
a top loop and a virtual singlet. By parametrising the coupling of the singlet to the electron with the $b_e$ parameter
in analogy to eq.~(\ref{eq:top_eta_coupling}), we find
\begin{equation}
\frac{d_e}{e} \simeq \frac{16}{3}\frac{\alpha_\textsc{em}}{(4\pi)^3} \frac{m_e}{f^2} {\rm Im}[b b_e^*] \left(\log \frac{m_t^2}{m_\eta^2} + 2\right)\,. 
\end{equation}
This leads to the bound
\begin{equation}
\frac{{\rm Im}[b b_e^*]}{f^2} \lesssim 4.5 \times 10^{-2} \,\, \textrm{TeV}^{-2}\,.
\end{equation}
This bound can obviously be evaded if the coupling of the singlet to the electron is small, or if the phases of the embedding
of the various elementary fermions are the same, so that $b$ and $b_e$ have the same complex phase. Notice that the latter condition is automatically realised if the embedding of the fermions preserves the $P_\eta$ symmetry, in which case all the phases are $\pm \pi$.
Analogous, although slightly weaker, bounds can be found on the couplings of the singlet with the light quarks, which
induce a contribution to the neutron EDM.

\section{Conclusions}
\label{sec:conclusions}

The dynamics of a strongly first-order EW phase transition, the structure of the gravitational wave spectrum
and the possibility to realise EW baryogenesis  can be successfully linked within the framework
of non-minimal composite Higgs models.
We illustrated this result by considering a scenario in which a Higgs sector extended with an extra singlet $\eta$
is described by NGBs associated to the strong-sector symmetry breaking pattern $\SO(6) \rightarrow \SO(5)$.
Within this scheme, a (possibly approximate) discrete $Z_2$ symmetry can be realised, which forbids a doublet-singlet mixing,
obtaining a viable form for the scalar potential and minimising the bounds from experimental constraints.

The scalar sector of this model is notoriously difficult to be tested directly at the LHC~\cite{Franzosi:2016aoo,Niehoff:2016zso,Banerjee:2017wmg}.
More accessible signatures could be provided by the interplay of the top partner dynamics with the additional singlet,
which clearly deserves further investigation (see ref.~\cite{Chala:2017xgc} for a model-independent analysis focused on
top partner searches).

Interestingly, future space-based gravitational interferometry  experiments could provide an alternative
way to test the dynamics of the EW phase transition in this class of models.
Even though it could be too ambitious to think that in the near future gravitational wave experiments will be competitive to collider searches in the study of the Higgs sector, it is more than plausible that the Higgs phenomenology will take advantage of new and complementary observational data.

Non-minimal composite Higgs scenarios with an extended Higgs sector have been already studied in the past, also in
connection to the EW phase transition, but only with a very narrow choice of the top partner representations. In the present work, for the first time, we consider a larger
set of representations, which are representative of a much wider spectrum of phenomenological signatures.
This variety opens up more interesting configurations for realising a strong first-order EW transition, which can be
obtained through a two-step process in which the singlet acquires a VEV at intermediate temperatures.
These set-ups can also open up the intriguing possibility of connecting EW symmetry breaking with baryogenesis.

We identified three main benchmark scenarios, depending on the embedding of the left-handed and right-handed
top components into $\SO(6)$ multiplets. The minimal set-up with the embedding $(q_L, t_R) \sim ({\bf 6}, {\bf 6})$
is the most constrained scenario. In this case stringent upper bounds on the singlet mass ($m_\eta < m_h/\sqrt{2}$),
and on the coupling of the portal interaction  $h^2\,\eta^2$ ($\lambda_{h\eta} < \lambda_h$) are found. A second scenario, the
$({\bf 15}, {\bf 6})$ model, partially relaxes these bounds to $m_\eta < m_h$ and $\lambda_{h\eta} < 2\lambda_h$.
Finally, the third benchmark, the $({\bf 6}, {\bf 20}')$ model, provides enough freedom in the structure of the scalar
potential to allow for (almost) arbitrary values of all the parameters.

Although a successful two-step phase transition with a strong EW symmetry breaking can be realised in all the benchmark
scenarios, a significant amount of correlation among the parameters of the models is needed in the $({\bf 6}, {\bf 6})$
and $({\bf 15}, {\bf 6})$ set-ups. A much larger viable region in the parameter space (and thus a smaller
amount of tuning) is instead available in the $({\bf 6}, {\bf 20}')$ model.
One of the crucial limitations in obtaining a strong first-order transition, although a two-minima structure
of the potential is quite common in these models, is the fact that a successful bubble nucleation is hard to satisfy.
This constraint is particularly severe for $m_\eta \lesssim 100$ GeV, which explains why the simpler benchmarks
offer a very narrow viable parameter space region.

We studied several important properties of the first-order phase transition, including the amount of supercooling,
the vacuum energy density, its time duration (important for the GW signals) and the strength of the phase transition
(crucial for the EW baryogenesis).
Interestingly, our estimates show that, in a significant fraction of the parameter space, the stochastic GW
background generated by the EW phase transition has peak frequencies within the sensitivity range of future space-based
experiments like LISA, BBO and DECIGO.
 This result is very interesting because in the same region of the parameter space EW baryogenesis could also be achievable.

As well known, EW baryogenesis requires additional sources of CP violation. In composite Higgs scenarios
the NGB nature of the Higgs naturally provides such ingredient.
 In particular, the presence of the non-renormalisable operator $\eta \, h \, \bar t_L  t_R $, which can have a complex coefficient and is naturally present in the $(\bf{6}, {\bf 6})$ and $(\bf{6}, {\bf 20}')$ scenarios, can trigger additional CP-violating effects.
 At $T = 0$, in the EWSB vacuum, this operator gives rise to small CP violating effects which can be compatible with
 the present constraints. On the other hand, when  both the Higgs and the singlet $\eta$ get a VEV, a new complex phase
 is induced in the top mass. This happens in the bubble walls during the EW phase transition.
For the $(\bf{6}, {\bf 6})$ and $(\bf{6}, {\bf 20}')$ models we found that a realistic baryon asymmetry can be generated
 for natural values of the parameters.
 
The results we obtained clearly show that non-minimal composite Higgs models offer a reach phenomenology
and deserve a careful investigation both at the theoretical and at the experimental level. There are a few aspects that  would be interesting to reconsider with a more  detailed study. One of these is the
GW signal, which we estimated by using previous results derived within a 2-Higgs doublet model set-up. Although we believe
that our analysis correctly captures the qualitative (and semi-quantitative) features of the GW spectrum, non-negligible
corrections could be present, which can be important to fully assess the detectability of the signal. Another aspect that
deserves a detailed study is EW baryogenesis. In sec.~\ref{sec:EW_baryo} we used some  estimates of the amount
of baryon asymmetry generated during the EW phase transition. A full analysis
taking into account the non-equilibrium dynamics by solving the Boltzmann equations is left for future work.

\section*{Note added}

During the completion of the manuscript, ref.~\cite{Bian:2019kmg} appeared, which presented a study of
the EW phase transition in $\SO(6)/\SO(5)$ CHMs. Ref.~\cite{Bian:2019kmg} focuses on the structure of the effective
potential in models with fermions in the $\bf 6$ and $\bf 15$ representations,
reaching conclusions similar to the ones we discussed in secs.~\ref{sec:(6,6)-model} and \ref{sec:(15,6)-model}.

\section*{Acnowledgments}

We thank Sebastian Bruggisser, Michele Redi, Andrea Tesi and Carroll L. Wainwright for useful discussions. LDR and GP would like to thank
the Galileo Galilei Institute for Theoretical Physics (GGI, Florence), for hospitality  at the final stage of this work during the research program on ``Next Frontiers in the Search for Dark Matter''. This research is supported in part by the MIUR under contract 2017FMJFMW (PRIN2017).

\appendix

\section{Structure of the effective potential}
\label{app:potential}

We collect here a few results regarding the structure of the effective potential. In the first subsection we focus on
the $P_\eta$ symmetry, discussing the conditions that ensure it to be preserved.
In the other subsections we report the explicit form of the fermion contributions to the effective potential
in the $({\bf 6}, {\bf 20})$ model and we discuss the gauge contributions to the Higgs potential.

\subsection{The $P_\eta$ invariance}\label{app:p_eta}

As we discussed in the main text, an important ingredient needed to obtain a viable structure for the effective potential
is the discrete $P_\eta$ invariance, which acts as $\eta \rightarrow - \eta$.
This symmetry guarantees that no tadpole term for the singlet is present in the
potential, allowing the standard EWSB point $(v, 0)$ to be a minimum.

As we anticipated in sec.~\ref{sec:(6,6)-model}, the $P_\eta$ invariance of the potential can be ensured by
a particular choice of the embedding phases, namely by fixing them to be $\pm \pi/2$,
provided that the composite dynamics is invariant under $P_\eta$ and CP. We will now prove this statement.

For definiteness we focus on the $({\bf 6}, {\bf 6})$ model, but similar considerations can be applied to the other set-ups.
In this model the $P_\eta$ invariance is obtained for $\alpha_{u_6} = \pm \pi/2$ for arbitrary values of the $\theta_{u_6}$
parameter. The mixing of the left-handed quark doublet with the composite dynamics (parametrised by the
$\Lambda_{q_L}^{\bf 6}$ spurion) is trivially invariant under $P_\eta$, so we only need to consider the
right-handed spurion $\Lambda_{u_R}^{\bf 6}$.
When $\alpha_{u_6} = \pm\pi/2$, the $P_\eta$ symmetry acts on $\Lambda_{u_R}^{\bf 6}$
as a kind of complex conjugation operation, namely it is equivalent to take the complex conjugate of the vector components
in eq.~(\ref{eq:ferm_spurions}) while keeping invariant the $\lambda_{u_R}$ prefactor.
We can extend this transformation to a more useful form by joining it with a CP transformation acting on the fermions
as $\psi \rightarrow \bar \psi$. The new transformation, which we denote by $\widetilde P_\eta$, corresponds to the
combination of ${\rm O}(5)$ and CP.
Under $\widetilde P_\eta$, the elementary--composite mixing terms are invariant up to the exchange $\lambda_\psi \rightarrow \lambda_\psi^*$.
If the composite sector is symmetric under $\widetilde P_\eta$,
then the terms that appear in the effective potential are invariant as well, apart from the replacement $\lambda_\psi \rightarrow \lambda_\psi^*$.
Due to the $\U(1)_X$ and $\SU(2)_L$ symmetries, however, in all the terms of the effective potential
the $\lambda$'s must always appear in the combination $\lambda_\psi \lambda_\psi^*$, which is invariant under
$\lambda_\psi \rightarrow \lambda_\psi^*$. For this reason the whole effective potential is also invariant under
$P_\eta$.

It is easy to check that, even if the composite sector is not invariant under $\widetilde P_\eta$, the lowest order terms in the
effective potential for embedding phases equal to $\pm\pi/2$ accidentally respect $P_\eta$.
In particular this is true for all invariants up to quartic order in the spurions that can arise at one-loop level
in the $({\bf 6}, {\bf 6})$ and $({\bf 15}, {\bf 6})$ models (for $\alpha_{u_6} =\pm \pi/2$ and $\alpha_{q_{15}} =\pm \pi/2$,
respectively). In the $({\bf 6}, {\bf 20}')$ model, instead, only the quadratic invariants accidentally respect $P_\eta$ for
$\alpha_{u_{20}} = \pm \pi/2$ and $\beta_{u_{20}} = \pm \pi/2$. The quartic terms, instead, can give rise to a breaking
of $P_\eta$.

\subsection{Fermions in the ($\bold{20}'$) representation}\label{app:fermions_20}

In this appendix we collect the explicit form of the fermion contributions to the effective potential
in the $({\bf 6}, {\bf 20}')$ model discussed in sec.~\ref{sec:rep_20}.
In this set-up three independent invariants can be built at the quadratic order in the spurions, namely
\begin{eqnarray}
{\cal O}^{(2)}_{q_L} &\equiv& \Sigma^T\!\cdot\!{\Lambda_{q_L}^{\bf 6}}\cdot{\Lambda_{q_L}^{\bf 6}}^\dagger\!\cdot\!\Sigma
= \frac{1}{2} |\lambda_{q_L}|^2 s_h^2\,,\\
{\cal O}^{(2, 1)}_{u_R} &\equiv& \Sigma^T\!\cdot\!{\Lambda_{u_R}^{\bf 20}}^\dagger\cdot\Lambda_{u_R}^{\bf 20}\!\cdot\!\Sigma
= \frac{1}{4} |\lambda_{u_R}|^2 f^{(1)}_{u_{20}}(h, \eta)\,,\\
{\cal O}^{(2, 2)}_{u_R} &\equiv& \left|\Sigma^T\!\cdot\Lambda_{u_R}^{\bf 20}\!\cdot\!\Sigma\right|^2
= |\lambda_{u_R}|^2 \left|f^{(2)}_{u_{20}}(h, \eta)\right|^2\,,
\end{eqnarray}
where
\bea
f^{(1)}_{u_{20}}(h, \eta) &=& \frac{1}{3}\cos^2 \theta_{u_{20}} + (1 + \sin^2 \theta_{u_{20}} ) c_h^2 + \frac{8}{\sqrt{6}} s_\eta \sqrt{1 - s_h^2 - s_\eta^2} \cos \alpha_{u_{20}}  \sin 2 \theta_{u_{20}}  \cos \phi_{u_{20}}\nn \\
&& - \frac{4}{\sqrt{6}} \left(1- s_h^2 - 2 s_\eta^2\right) \cos (\alpha_{u_{20}} - \beta_{u_{20}})  \sin 2 \theta_{u_{20}} \sin \phi_{u_{20}} \,, \nn \\
f^{(2)}_{u_{20}}(h, \eta) &=&  \frac{1}{4 \sqrt{3}} e^{i \alpha_{u_{20}}} \cos \theta_{u_{20}} (1 + 3 c_{2h}) \\
&& + \sin \theta_{u_{20}} \left(\sqrt{2} s_\eta \sqrt{1 - s_h^2 - s_\eta^2} \cos \phi_{u_{20}} + \frac{1}{\sqrt{2}} e^{i \beta{u_{20}}} (s_h^2 + 2 s_\eta^2 - 1) \sin \phi_{u_{20}} \right) \,.\nn
\eea

Several invariants can be built at the quartic level. There is only one involving the left-handed spurion, which obviously
coincides with the one we found in the $({\bf 6}, {\bf 6})$ model:
\begin{equation}
{\cal O}^{(4)}_{q_L} \equiv (\Sigma^T\cdot{\Lambda_{q_L}^{\bf 6}}\cdot{\Lambda_{q_L}^{\bf 6}}^\dagger\cdot\Sigma)^2
= \frac{1}{4} |\lambda_{q_L}|^4 s_h^4\,.
\end{equation}
Three independent invariants can be constructed with $\Lambda_{u_R}^{\bf 20}$:
\begin{eqnarray}
{\cal O}^{(4, 1)}_{u_R} &\equiv& \left(\Sigma^T\!\cdot\!{\Lambda_{u_R}^{\bf 20}}^\dagger\cdot\Lambda_{u_R}^{\bf 20}\!\cdot\!\Sigma\right)^2
= \frac{1}{16} |\lambda_{u_R}|^4 \left(f^{(1)}_{u_{20}}(h, \eta)\right)^2\,,\label{eq:O41uR20}\\
{\cal O}^{(4, 2)}_{u_R} &\equiv& \left|\Sigma^T\!\cdot\Lambda_{u_R}^{\bf 20}\!\cdot\!\Sigma\right|^4
= |\lambda_{u_R}|^4 \left|f^{(2)}_{u_{20}}(h, \eta)\right|^4\,,\label{eq:O42uR20}\\
{\cal O}^{(4, 3)}_{u_R} &\equiv& \left(\Sigma^T\!\cdot\!{\Lambda_{u_R}^{\bf 20}}^\dagger\cdot\Lambda_{u_R}^{\bf 20}\!\cdot\!\Sigma\right) \left|\Sigma^T\!\cdot\Lambda_{u_R}^{\bf 20}\!\cdot\!\Sigma\right|^2
= \frac{1}{4} |\lambda_{u_R}|^4 f^{(1)}_{u_{20}}(h, \eta) \left|f^{(2)}_{u_{20}}(h, \eta)\right|^2\,.\label{eq:O43uR20}
\end{eqnarray}
Finally, four left-right invariants are present:
\begin{eqnarray}
{\cal O}^{(4, 1)}_{q_L u_R} &\equiv& \Sigma^T\!\cdot\!{\Lambda_{u_R}^{\bf 20}}^\dagger\!\cdot\!\Lambda_{q_L}^{\bf 6}\!\cdot\!{\Lambda_{q_L}^{\bf 6}}^\dagger\!\cdot\!{\Lambda_{u_R}^{\bf 20}}\!\cdot\!\Sigma
= \frac{1}{24} |\lambda_{q_L} \lambda_{u_R}|^2 \cos^2 \theta_{u_{20}}\, s_h^2\,,\label{eq:O41qLuR20}\\
{\cal O}^{(4, 2)}_{q_L u_R} &\equiv& \left|\Sigma^T\!\cdot\!\Lambda_{u_R}^{\bf 20}\!\cdot\!\Sigma\right|^2 \left(\Sigma^T\!\cdot\!{\Lambda_{q_L}^{\bf 6}}\!\cdot\!{\Lambda_{q_L}^{\bf 6}}^\dagger\!\cdot\!\Sigma\right)
= \frac{1}{2} |\lambda_{q_L} \lambda_{u_R}|^2 s_h^2 \left|f^{(2)}_{u_{20}}(h, \eta)\right|^2\,,\label{eq:O42qLuR20}\\
{\cal O}^{(4, 3)}_{q_L u_R} &\equiv& {\rm Re}\left[\Sigma^T\!\cdot\!{\Lambda_{u_R}^{\bf 20}}^\dagger\!\cdot\!\Lambda_{q_L}^{\bf 6}\!\cdot\!{\Lambda_{q_L}^{\bf 6}}^\dagger\!\cdot\!\Sigma \;\,\Sigma^T\!\cdot\!\Lambda_{u_R}^{\bf 20}\!\cdot\!\Sigma\right]\nn\\
&=& -\frac{1}{4\sqrt{3}} |\lambda_{q_L} \lambda_{u_R}|^2 \cos \theta_{u_{20}} s_h^2\; {\rm Re}\left[e^{-i \alpha_{u_{20}}} f^{(2)}_{u_{20}}(h, \eta)\right]\,,\label{eq:O43qLuR20}\\
{\cal O}^{(4, 4)}_{q_L u_R} &\equiv& {\rm Im}\left[\Sigma^T\!\cdot\!{\Lambda_{u_R}^{\bf 20}}^\dagger\!\cdot\!\Lambda_{q_L}^{\bf 6}\!\cdot\!{\Lambda_{q_L}^{\bf 6}}^\dagger\!\cdot\!\Sigma \;\,\Sigma^T\!\cdot\!\Lambda_{u_R}^{\bf 20}\!\cdot\!\Sigma\right]\nn\\
&=& -\frac{1}{4\sqrt{3}} |\lambda_{q_L} \lambda_{u_R}|^2 \cos \theta_{u_{20}} s_h^2\; {\rm Im}\left[e^{-i \alpha_{u_{20}}} f^{(2)}_{u_{20}}(h, \eta)\right]\,.\label{eq:O44qLuR20}
\end{eqnarray}
Notice that the ${\cal O}^{(4, 1)}_{q_L u_R}$ does not really give rise to a new structure in the potential since its
dependence on the scalar fields coincides with the one of the quadratic invariant ${\cal O}^{(2)}_{q_L}$. One can thus
neglect this invariant in the study of the potential.

Regarding the $P_\eta$ symmetry a difference with respect to the $({\bf 6}, {\bf 6})$ and $({\bf 15}, {\bf 6})$ is present.
The conditions on the phases of the embedding, namely $\alpha_{u_{20}} = \pm \pi/2$ and $\beta_{u_{20}} = \pm \pi/2$,
are not enough to ensure that the invariants up to quartic order respect $P_\eta$.
The ${\cal O}^{(4, 4)}_{q_L u_R}$ invariant, indeed, breaks the discrete symmetry and generates an interaction
of the form $h^2 \eta$. To ensure an unbroken $P_\eta$, one needs to also assume that the composite sector is
invariant under CP, in which case the ${\cal O}^{(4, 4)}_{q_L u_R}$ invariant is forbidden.

\begin{table}
\centering
\small
\begin{tabular}{@{\,}c@{\;}||@{\;}c@{\;}|@{\;}c@{\;}|@{\;}c@{\,}}
$({\bf 6}, {\bf 20}')$ & ${\cal O}_{q_L}^{(2)}$ & ${\cal O}_{u_R}^{(2,1)}$ & ${\cal O}_{u_R}^{(2,2)}$\\
\hline\hline
\rule[-8pt]{0pt}{2.25em}$\frac{N_c}{16 \pi^2}\times$ & $\lambda_{q_L}^2 \frac{m_\Psi^2}{f^2}$ & $\lambda_{u_R}^2 \frac{m_\Psi^2}{f^2}$ & $\lambda_{u_R}^2 \frac{m_\Psi^2}{f^2}$\\
\hline\hline
\rule[-7pt]{0pt}{1.75em}$\mu_h^2/f^2$ & $1$ & $\frac{1}{2} (1+ s_\theta^2) - \frac{2}{\sqrt{6}} s_{2 \theta} s_\phi$ & $ \frac{5}{\sqrt{6}} s_{2 \theta} s_\phi - 2 (1 - s_\theta^2 c_\phi^2)$\\
\rule[-7pt]{0pt}{2em}$\mu_\eta^2/f^2$ & $0$ & $\frac{4}{\sqrt{6}} s_{2 \theta} s_\phi$ & $4 s_\theta^2 c_{2 \phi} + \frac{4}{\sqrt{6}} s_{2 \theta} s_\phi$\\
\hline
\rule[-7pt]{0pt}{1.75em}$\lambda_h$ & $-\frac{2}{3}$ & $\frac{4}{3 \sqrt{6}} s_{2 \theta}s_\phi - \frac{1}{3} ( 1+ s_\theta^2)$ & $\frac{39}{8} c_\theta^2 - \frac{14\sqrt{6}}{9} s_{2 \theta} s_\phi + \frac{10}{3} s_\theta^2 s_\phi^2$\\
\rule[-7pt]{0pt}{2em}$\lambda_\eta$ & $0$ & $-\frac{8}{3\sqrt{6}} s_{2 \theta} s_\phi$ & $- \frac{32}{3} c_{2 \phi} s_\theta^2 - \frac{8}{3 \sqrt{6}} s_{2 \theta} s_\phi$\\
\rule[-7pt]{0pt}{2em}$\lambda_{h\eta}$ & $0$ & $0$ & $- 4 c_{2 \phi} s_\theta^2 - \sqrt{6} s_{2 \theta} s_\phi$
\end{tabular}\\
\vspace{1.35em}
\begin{tabular}{@{\,}c@{\;}||@{\;}c@{\;}|@{\;}c@{\;}|@{\;}c@{\,}}
$({\bf 6}, {\bf 20}')$ & ${\cal O}_{q_L}^{(4)}$ & $\widetilde {\cal O}_{q_L u_R}^{(4,2)}$ & $\widetilde {\cal O}_{q_L u_R}^{(4,3)}$\\
\hline\hline
\rule[-8pt]{0pt}{2.em}$\frac{N_c}{16 \pi^2}\times$ & $\lambda_{q_L}^4$ & $\lambda_{q_L}^2 \lambda_{u_R}^2$ & $\lambda_{q_L}^2 \lambda_{u_R}^2$\\
\hline\hline
\rule[-7pt]{0pt}{2em}$\lambda_h$ & $1$ & $ \frac{5}{\sqrt{6}} s_{2 \theta} s_\phi - 2 (1 - s_\theta^2 c_\phi^2)$ & $\frac{1}{2} c_\theta^2- \frac{1}{2\sqrt{6}}s_{2\theta} s_\phi$\\
\rule[-7pt]{0pt}{1.75em}$\lambda_\eta$ & $0$ & $0$ & $0$\\
\rule[-7pt]{0pt}{1.75em}$\lambda_{h\eta}$ & $0$ & $2 s_\theta^2 c_{2\phi} + \frac{2}{\sqrt{6}} s_{2 \theta} s_\phi$ & $- \frac{1}{2\sqrt{6}} s_{2 \theta} s_\phi$
\end{tabular}\\
\vspace{1.35em}
\begin{tabular}{@{\,}c@{\;}||@{\;}c@{\;}|@{\;}c@{\,}}
$({\bf 6}, {\bf 20}')$ & $\widetilde {\cal O}_{u_R}^{(4,1)}$ & $\widetilde {\cal O}_{u_R}^{(4,2)}$\\
\hline\hline
\rule[-8pt]{0pt}{2em}$\frac{N_c}{16 \pi^2}\times$ & $\lambda_{u_R}^4$ & $\lambda_{u_R}^4$\\
\hline\hline
\rule[-7pt]{0pt}{2.25em}$\lambda_h$ & $\left(\frac{2}{\sqrt{6}} s_{2\theta} s_\phi -\frac{1}{2}(1+s_\theta^2)\right)^2$ & $\left(\frac{5}{\sqrt{6}} s_{2 \theta} s_\phi - 2 (1 - s_\theta^2 c_\phi^2)\right)^2$\\
\rule[-7pt]{0pt}{2.25em}$\lambda_\eta$ & $\frac{8}{3} s_{2\theta}^2 s_\phi^2$ & $\left(4 s_\theta^2 c_{2 \phi} + \frac{4}{\sqrt{6}} s_{2 \theta} s_\phi\right)^2$\\
\rule[-7pt]{0pt}{2em}$\lambda_{h\eta}$ & $\frac{4}{\sqrt{6}} s_{2\theta} s_\phi \left(\frac{2}{\sqrt{6}} s_{2\theta} s_\phi -\frac{1}{2}(1+s_\theta^2)\right)$ & $\left(\frac{5}{\sqrt{6}} s_{2 \theta} s_\phi - 2 (1 - s_\theta^2 c_\phi^2)\right)\left(4 s_\theta^2 c_{2 \phi} + \frac{4}{\sqrt{6}} s_{2 \theta} s_\phi\right)$
\end{tabular}\\
\vspace{1.35em}
\begin{tabular}{@{\,}c@{\;}||@{\;}c@{\,}}
$({\bf 6}, {\bf 20}')$ & $\widetilde {\cal O}_{u_R}^{(4,3)}$\\
\hline\hline
\rule[-8pt]{0pt}{2.em}$\frac{N_c}{16 \pi^2}\times$ & $\lambda_{u_R}^4$\\
\hline\hline
\rule[-7pt]{0pt}{2em}$\lambda_h$ & $\left(\frac{2}{\sqrt{6}} s_{2\theta} s_\phi -\frac{1}{2}(1+s_\theta^2)\right)\left(\frac{5}{\sqrt{6}} s_{2 \theta} s_\phi - 2 (1 - s_\theta^2 c_\phi^2)\right)$\\
\rule[-7pt]{0pt}{2em}$\lambda_\eta$ & $\frac{4}{\sqrt{6}} s_{2\theta} s_\phi\left(4 s_\theta^2 c_{2 \phi} + \frac{4}{\sqrt{6}} s_{2 \theta} s_\phi\right)$\\
\rule[-7pt]{0pt}{2em}$\lambda_{h\eta}$ & $\frac{2}{\sqrt{6}} s_{2\theta} s_\phi \left(\frac{5}{\sqrt{6}} s_{2 \theta} s_\phi - 2 (1 - s_\theta^2 c_\phi^2)\right) + \left(2 s_\theta^2 c_{2 \phi} + \frac{2}{\sqrt{6}} s_{2 \theta} s_\phi\right)\left(\frac{2}{\sqrt{6}} s_{2\theta} s_\phi -\frac{1}{2}(1+s_\theta^2)\right)$
\end{tabular}
\caption{Contributions to the coefficients of the effective potential coming from the quadratic and quartic one-loop invariants
from fermions in the $({\bf 6}, {\bf 20}')$ model. The results are given for the $P_\eta$-symmetric case $\alpha_{u_{20}} = \beta_{u_{20}} = \pm\pi/2$. The choice $\alpha_{u_{20}} = -\beta_{u_{20}} = \pm\pi/2$ can be obtained through the replacement $\phi \rightarrow -\phi$. For shortness we used the notation $s_\theta \equiv \sin \theta_{u_{20}}$, $c_\theta \equiv \cos \theta_{u_{20}}$ and analogously for $\phi_{u_{20}}$.
The second line reports the estimate of the size of the coefficients multiplying each invariant.
Notice that all the quartic invariants given in the last three tables have $\mu_h = \mu_\eta = 0$.
The definition of the $\widetilde{\cal O}$ operators is given in eqs.~(\ref{eq:OqLuR42_tilde})--(\ref{eq:OuR43_tilde}).\label{tab:invariants_20}}
\end{table}

In order to simplify the results a shifted form of the purely right and the mixed left-right invariants at the quartic level
has been used. The definitions are given by
\begin{eqnarray}
\widetilde {\cal O}_{q_L u_R}^{(4,2)} &\equiv& {\cal O}_{q_L u_R}^{(4,2)}  - \frac{|\lambda_{u_R}|^2}{6}
\left(2 \cos^2 \theta - \sqrt{6} \sin 2 \theta \sin \phi + 3 \sin^2 \theta \sin^2 \phi\right) {\cal O}_{q_L}^{(2)}\,,\label{eq:OqLuR42_tilde}\\
\widetilde {\cal O}_{q_L u_R}^{(4,3)} &\equiv& {\cal O}_{q_L u_R}^{(4,3)} - \frac{|\lambda_{u_R}|^2}{12}
\left(-2 \cos^2 \theta + \frac{\sqrt{6}}{2} \sin 2 \theta \sin \phi\right) {\cal O}_{q_L}^{(2)}\,,\label{eq:OqLuR43_tilde}
\end{eqnarray}
and
\begin{eqnarray}
\widetilde {\cal O}_{u_R}^{(4,1)} &\equiv& {\cal O}_{u_R}^{(4,1)} - \frac{|\lambda_{u_R}|^2}{3}
\left(2 + \sin^2 \theta - \sqrt{6} \sin 2 \theta \sin \phi\right) {\cal O}_{u_R}^{(2,1)}\,,\label{eq:OuR41_tilde}\\
\widetilde {\cal O}_{u_R}^{(4,2)} &\equiv& {\cal O}_{u_R}^{(4,2)} - \frac{|\lambda_{u_R}|^2}{6}
\left(2 \cos \theta - \sqrt{6} \sin \theta \sin \phi\right)^2 {\cal O}_{u_R}^{(2,2)}\,,\label{eq:OuR42_tilde}\\
\widetilde {\cal O}_{u_R}^{(4,3)} &\equiv& {\cal O}_{u_R}^{(4,3)}
- \frac{|\lambda_{u_R}|^2}{6} \left(2 + \sin^2 \theta - \sqrt{6} \sin 2 \theta \sin \phi\right) {\cal O}_{u_R}^{(2,1)}\nn\\
&&-\;\frac{|\lambda_{u_R}|^2}{12} \left(2 \cos \theta - \sqrt{6} \sin \theta \sin \phi\right)^2 {\cal O}_{u_R}^{(2,2)}\,.\label{eq:OuR43_tilde}
\end{eqnarray}
In order to write more compact formulae we used the simplified notation $\theta \equiv \theta_{u_{20}}$
and $\phi \equiv \phi_{u_{20}}$.

The coefficients of the effective potential coming from the quadratic and quartic invariants are listed
in table~\ref{tab:invariants_20} for the $P_\eta$-symmetric case, namely $\alpha_{u_{20}} = \pm \pi/2$
and $\beta_{u_{20}} = \pm \pi/2$. Notice that the ${\cal O}_{q_L u_R}^{(4,1)}$ invariant is not included since its contributions
have the same structure as the ones from ${\cal O}_{q_L}^{(2)}$.

\subsection{Gauge contributions}\label{app:gauge}

Since the SM gauge interactions break explicitly the global symmetry down to $\SU(2)_L \times \U(1)_Y \times \SO(2)_\eta$, the gauge loops generate a potential for the Higgs but not for the gauge singlet $\eta$.
The gauge contributions to the effective potential are typically neglected because they are parametrically smaller than the ones coming from the top quark sector.

To derive the form of the spurions~\cite{Panico:2011pw,DeSimone:2012fs,Panico:2015jxa} we need to specify the embedding of the
elementary EW gauge fields into the adjoint representation of the composite-sector global symmetry.
The $\SU(2)_L$ gauge fields $W_\mu^a$ gauge the corresponding subgroup of the unbroken $\SO(5)$. The $\U(1)_Y$
hypercharge boson $B_\mu$, instead gauges the combination $Y = T^3_R + X$, where $T_R^3$ is the third generator of the
$\SU(2)_R$ subgroup of $\SO(5)$, while $X$ corresponds to the generator of an additional (unbroken) $\U(1)_X$
global symmetry needed to accommodate the correct hypercharges for the fermions.

The spurions can be introduced by formally uplifting the gauge interactions to fully $\SO(6)$ invariant operators, namely
\begin{equation}
{\cal L}_{\textit gauge} = g W_\mu^a J_L^{\mu,a} + g' B_\mu J_R^{\mu,3} + g' B_\mu J_X^\mu\,,
\end{equation}
where $ J_L^{\mu,a}$, $J_R^{\mu,3}$ and $J_X^\mu$ denote the relevant composite-sector currents. Because the
$\U(1)_X$ current is a singlet under $\SO(6)$ it does not generate any potential for the Higgses, so we can focus on the
remaining two terms. We can rewrite them as
\begin{equation}
{\cal L}_{\textit gauge} = {\cal G}^a_I W_\mu^a J^{\mu,I} + {\cal G}'_I B_\mu J^{\mu,I}\,,
\end{equation}
where the index $I$ runs over the $15$ components of the $\SO(6)$ current multiplet $J^{\mu,I}$. The two spurions,
which we can rewrite in matrix notation as
\begin{equation}
{\cal G} = {\cal G}^a_I T^I\,, \qquad \quad {\cal G}' = {\cal G}'_I T^I\,,
\end{equation}
formally transform in the adjoint representation of $\SO(6)$, making the interaction terms fully invariant.

In order to derive the contributions to the effective potential we must construct invariant terms using the spurions
and the Goldstone multiplet $\Sigma$. At the leading order in an expansion in $g, g'$, we find two invariants:
\begin{eqnarray}
O_{g^2} &=& \sum_a \Sigma^T\cdot{\cal G}^a\cdot{\cal G}^a\cdot\Sigma = \frac{3}{4} g^2 \sin^2 \frac{h}{f}\,,\\
O_{g'^2} &=& \Sigma^T\cdot{\cal G}'\cdot{\cal G}'\cdot\Sigma = \frac{1}{4} g'^2 \sin^2 \frac{h}{f}\,.
\end{eqnarray}
The leading gauge contribution to the effective potential is thus given by\footnote{Due to the symmetry among the
various gauge components with respect to $\SO(5)$, each gauge degree of freedom contributes in the same
way to the potential. For this reason the coefficient $c_g$ in front of the two invariants is the same.}
\begin{equation}
V^{(2)}_{\textit gauge} = \frac{g_\rho^2}{16 \pi^2} f^4 c_g \left(\frac{3}{4} g^2 + \frac{1}{4} g'^2\right) \sin^2 \frac{h}{f}\,,
\end{equation}
where $c_g$ is a coefficient expected to be of order one. This potential corresponds to the following contributions to the
coefficients of the expanded effective potential
\begin{equation}
\mu_h^2 = \frac{g_\rho^2}{16 \pi^2} f^2 c_g \left(\frac{3}{2} g^2 + \frac{1}{2} g'^2\right)\,,
\qquad
\lambda_h = -\frac{g_\rho^2}{16 \pi^2} c_g \left(g^2 + \frac{1}{3} g'^2\right)\,.
\end{equation}
As can be seen from the above expressions, the gauge contribution to the
effective potential depends only on $h$ and not $\eta$. This property is a consequence of the fact that the coupling
of the elementary gauge fields with the composite sector does not break the $\SO(2)_\eta$ subgroup.

\section{The thin-wall approximation}
\label{app:thinwall}

It is instructive to discuss the thin-wall limit as the analytical approximations can help to gain some insights into the features of the phase transition.
In this regime, the $r$-dependent friction term in eq.~(\ref{eq:bounce}) can be neglected and the bounce solution with $\epsilon \to 0$ is implicitly defined as
\bea
r = \int_\phi^{\phi_T} \frac{d \varphi}{\sqrt{V(\varphi, T)}}
\eea
where the integral is evaluated along the path that extremises the action. While the single-field problem is trivial, in the multi-field case the direction of the tunnelling path in field-space is not known a priori. 
A good approximation is obtained by evaluating the integral along the path that ``minimises'' the potential.\footnote{The
concept of ``minimising'' the potential away from a local or global minimum is intrinsically ambiguous. To obtain a reasonable
approximation, we constructed a curve that connects the true and false vacua by minimizing the potential in the $h$
direction at fixed $\eta$. For our potential this prescription leads to numerical results in fair agreement with the exact ones.} Using the potential in the high-temperature expansion as defined in eq.~(\ref{eq:V_highT}) and around the critical temperature, the bounce configuration is explicitly given by
\bea
\label{eq:bounce_2}
h(r) = v_c \, \textrm{sech} \frac{r}{L_w} \,, \qquad \textrm{with} \quad L_w^{2} v_c^2 = \frac{2}{\bar \lambda} \frac{\lambda_\eta}{\lambda_h}  \,, \quad \bar \lambda = \lambda_{h\eta} - \sqrt{\lambda_h \lambda_\eta} 
\eea
where $\eta^2 = w_c^2(1 - h^2/v_c^2)$ and $L_w$ represent the thickness of the bubble wall.

The euclidean action evaluated on the bubble configuration in the thin-wall limit is $S_3 = 16 \pi S_1^3/(3 \epsilon^2)$ with $\epsilon = V(\phi_T, T) - V(\phi_F, T)$.
It originates from the minimisation of two competing contributions: the volume energy $-4/3 \pi R^3 \epsilon$, with $R$ being the bubble radius, and the surface energy $4 \pi R^2 S_1$, where $S_1$ corresponds to the one-dimensional action, which in our case reads as
\bea
S_1 = \frac{1}{3} v_c^3 \left( \frac{\lambda_h}{\lambda_\eta} \right)^{\frac{1}{4}} \sqrt{\bar \lambda} \,.
\eea
The three-dimensional action is then given by
\bea
\label{eq:d3action}
\frac{S_3}{T} = \left( \frac{64 \pi}{81} \right) \frac{v_c^5}{T(T^2- T_c^2)^2}  \gamma  
= \left( \frac{16 \pi}{81} \right) \left( \frac{v_c}{T_c} \right)^5   \frac{\gamma}{\Delta x^2} + \mathcal O(\Delta x^{-1})
\eea
with
\bea
\gamma =  \left( \frac{\lambda_h}{\lambda_\eta} \right)^{3/4} \frac{\bar \lambda^{3/2}}{\left( c_h - c_\eta \sqrt{\lambda_h/\lambda_\eta}\right)^2} \,,
\eea
and $c_h, c_\eta$ defined in eq.~(\ref{eq:ch_ceta}).
In the last equality of eq.~(\ref{eq:d3action}) we expanded $S_3/T$ around the critical temperature and expressed it in terms of 
$\Delta x = (T_c - T)/T_c \ll 1$ which parametrises the amount of supercooling. As we already mentioned, the latter can be estimated by determining the nucleation temperature $T_n$, through the condition $(S_3/T)|_{T = T_n} = 140$, which is  given by
\bea
T_n \simeq  T_c \left[ 1 - \frac{2}{9} \sqrt{\frac{\pi}{35}}   \left( \frac{v_c}{T_c} \right)^{5/2} \sqrt{\gamma}  \right] \,.
\eea
The estimate above is typically within $\sim 20 \%$ from the result obtained from the numerical simulations for $v_n/T_n \lesssim 1.5$. The agreement degrades, as expected, for stronger phase transitions. Moreover, for fixed $\lambda_{h \eta}$, the approximation improves for increasing $\lambda_\eta$.

Another key parameter characterising the phase transition is its inverse time duration, normalised to the Hubble rate, that can be computed from the temperature variation of the action. 
In the same thin-wall limit, this reads as
\bea
\frac{\beta}{H_n} = T \frac{d}{dT} \frac{S_3}{T} \bigg|_{T_n} = 1260 \sqrt{\frac{35}{\pi}}   \left( \frac{v_c}{T_c} \right)^{-5/2} \gamma^{-1/2}
\eea 
where the numerical factor, $\sim 4200$, sets the typical scale of the $\beta/H_n$ parameter. Notice that, a stronger phase transition corresponds to a longer nucleation. The approximate analytic result is $\sim 40 \%$ smaller than the numerical one for $v_n/T_n \lesssim 2$ and $\lambda_\eta \lesssim 2$. The discrepancy reduces below $\sim 20 \%$ for larger $\lambda_\eta$. 

Finally, the vacuum energy density released during the phase transition, in the weak supercooling regime, is 
\bea
\label{eq:rho_approx}
\rho_\textrm{vac} = \frac{T_c^4}{2} \left( \frac{v_c}{T_c} \right)^2  \left( c_h - c_\eta \sqrt{\lambda_h/\lambda_\eta}\right) + T_c^4 \left( \frac{c_h^2}{\lambda_h} - \frac{c_\eta^2}{\lambda_\eta} \right) \Delta x + \mathcal O(\Delta x^2)\,,
\eea 
while the latent heat reads $\ell = 2 \rho_\textrm{vac} (1 - \Delta x) + \mathcal O(\Delta x^2)$.  From eq.~(\ref{eq:rho_approx}) one can compute $\alpha$, the normalised vacuum energy density 
\bea
\alpha = \frac{\rho_\textrm{vac}}{\rho_\textrm{rad}} = \frac{15}{g_* \pi^2} \left( \frac{v_c}{T_c} \right)^2  \left( c_h - c_\eta \sqrt{\lambda_h/\lambda_\eta}\right)  + \mathcal O(\Delta x)\,,\label{eq:alpha_vc}
\eea
with $\rho_\textrm{rad} = g_* \pi^2 T^4/30$, where $g_*$ is the number of relativistic degrees of freedom at the temperature $T$. The analytic approximation for $\alpha$ deviates $\sim 10 \%$ from the numerical result for $v_n/T_n \lesssim 2$.

\providecommand{\href}[2]{#2}\begingroup\raggedright\endgroup

\end{document}